\documentclass[journal]{IEEEtran}
\usepackage{caption}
\usepackage{algorithmic}
\usepackage[linesnumbered,ruled]{algorithm2e}
\usepackage{graphics,epstopdf}
\usepackage{graphicx}
\usepackage{balance}
\usepackage{listings}
\usepackage{comment}
\usepackage{enumitem}
\usepackage{color}
\usepackage{fixmath}
\usepackage{cite}
\usepackage{subcaption}
\usepackage{amsmath,amssymb,amsfonts}
\usepackage{graphicx}
\usepackage{textcomp}
\usepackage{tabularx}
\usepackage{booktabs}
\usepackage{longtable}
\usepackage{multirow}
\usepackage{colortbl}
\usepackage{hhline}
\usepackage{tabularx,colortbl}
\usepackage{stmaryrd}
\usepackage{float}
\usepackage{cite}
\usepackage{enumitem}
\usepackage{soul}
\usepackage{bm} %bold math \bm{s}
\usepackage{mathtools} % text on arrow
\usepackage{mathrsfs} % for A as adversary
\usepackage{multirow}
\usepackage{array}

\usepackage{graphicx}
\usepackage{float}
\usepackage{caption}
\usepackage{lipsum}
\usepackage{fancyhdr}

\ifCLASSINFOpdf

\else

\fi

\definecolor{lightgray}{gray}{0.9}
\definecolor{g}{rgb}{0.56, 0.93, 0.56}

\definecolor{bb}{rgb}{0.56, 0.86, 0.93}

\definecolor{bb}{rgb}{1, 0.65, 0.79}

\newcommand{\etal}{{\lowercase{\textit{et al. }}}}

\begin{document}

%\title{Privacy-Preserving Traffic Forecasting System Using Functional Encryption and Deep Learning}
\title{Leveraging Functional Encryption and Deep Learning for Privacy-Preserving Traffic Forecasting}

\author{Isaac Adom, \IEEEmembership{Student Member, IEEE}, Mohammmad Iqbal Hossain, Hassan Mahmoud, Ahmad Alsharif, \IEEEmembership{Senior Member, IEEE}, Mahmoud Nabil Mahmoud, \IEEEmembership{Member, IEEE}, and Yang Xiao, \IEEEmembership{Fellow, IEEE} % \vspace{-3mm}
\thanks{M. I. Hossain, I. Adom, and M. N. Mahmoud are with Department of Electrical and Computer Engineering, North Carolina A\&T University, Greensboro, NC, USA. (e-mail:  mhossain1@aggies.ncat.edu; iadom@aggies.ncat.edu; mnmahmoud@ncat.edu).}
\thanks{H. Mahmoud, A. Alsharif, and Y. Xiao are with the Department of Computer Science, The University of Alabama, Tuscaloosa, AL, USA. (e-mail: hmahmoud@crimson.ua.edu; aalsharif@ieee.org; yangxiao@ieee.org).}
}
% The paper headers
%\markboth{\ Privacy-Preserving Traffic Forecasting System Using Deep Learning}%{}
%\markboth{IEEE Transactions on Intelligent Transportation Systems}{}%
%\markboth{M. I. Hossain \etal Privacy-Preserving Traffic Forecasting System Using Deep Learning}{}%

\markboth{I. Adom \etal Leveraging Functional Encryption and Deep Learning for Privacy-Preserving Traffic Forecasting}{}

\maketitle
%\maketitle
%\IEEEdisplaynontitleabstractindextext

%\IEEEpeerreviewmaketitle

%\IEEEtitleabstractindextext{%
\begin{abstract} Over the past few years, traffic congestion has continuously plagued the nation’s transportation system creating several negative impacts including longer travel times, increased pollution rates, and higher collision risks. To overcome these challenges, Intelligent Transportation Systems (ITS) aim to improve mobility and vehicular systems, ensuring higher levels of safety by utilizing cutting-edge technologies, sophisticated sensing capabilities, and innovative algorithms. Drivers’ participatory sensing, current/future location reporting, and machine learning algorithms have considerably improved real-time congestion monitoring and future traffic management. However, each driver’s sensitive spatiotemporal location information can create serious privacy concerns. To address these challenges, we propose in this paper a secure, privacy-preserving location reporting and traffic forecasting system that guarantees privacy protection of driver data while maintaining high traffic forecasting accuracy. Our novel k-anonymity scheme utilizes functional encryption to aggregate encrypted location information submitted by drivers while ensuring the privacy of driver location data. Additionally, using the aggregated encrypted location information as input, this research proposes a deep learning model that incorporates a Convolutional-Long Short-Term Memory (Conv-LSTM) module to capture spatial and short-term temporal features and a Bidirectional Long Short-Term Memory (Bi-LSTM) module to recover long-term periodic patterns for traffic forecasting. With extensive evaluation on real datasets, we demonstrate the effectiveness of the proposed scheme with less than 10\% mean absolute error for a 60-minute forecasting horizon, all while protecting driver privacy.

\end{abstract}

\begin{IEEEkeywords}

Privacy-preserving, functional encryption, deep learning, traffic forecasting. %\vspace{-0.4cm}
\end{IEEEkeywords}
%}

%\titlepgskip=-15pt

{\section{Introduction}\label{sec:introduction}}

%%%%%%%%%%%%%%%%%%%%%%%%%%%%%%%%%%%%%%%%%%%%%%%%%%%%%%%%

%%%%%%%%%%%%%%%%%%%%%%%%%%%%%%%%%%%%%%%%%%%%%%%%%%%%%%%%
\IEEEPARstart{T}{raffic} congestion in the transportation system substantially negatively impacts productivity, living standards, the economy, and the environment. It leads to financial losses, increased travel time, fuel consumption, operational costs, and environmental degradation because of excessive carbon dioxide emissions~\mbox{\cite{hansen2001determination}}. According to the Inrix location-based data analytic report\mbox{\cite{inrix_2023}}, Chicago and Boston were the top-ranked cities in terms of time and financial losses due to traffic congestion in 2023, with drivers in each city wasting up to 155 and 134 hours in traffic, respectively. Furthermore, Chicago drivers lose up to \$2,618 per year, while Boston drivers waste up to \$2,270, and similar statistics are seen in other cities \mbox{\cite{inrix_2023}}. Thus, governments are investing huge amounts of money to help reduce traffic congestion. If effective measures are not implemented promptly, the problem is anticipated to intensify, given the inability of traditional anti-congestion strategies to keep pace with the rapid economic growth of major cities.

\begin{figure}[t]
\centering
\includegraphics[width=0.95\columnwidth]{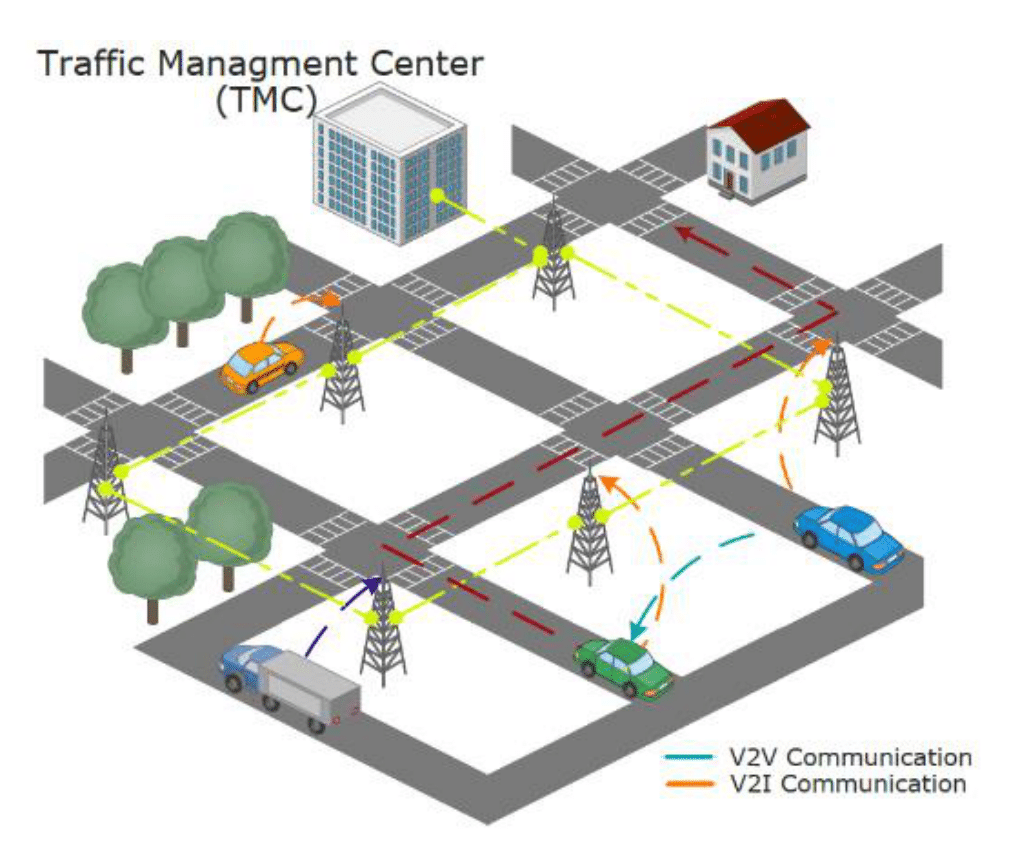}
\caption{Conceptual framework of Vehicular Ad-hoc Networks. \label{fig:SG_arch}}
\vspace{-0.3cm}
\end{figure}

To address these challenges, Intelligent Transportation Systems (ITS) play a pivotal role in mitigating the adverse effects of traffic congestion \mbox{\cite{Sharma2022Awasthi}}. One notable facet of ITS is developing and implementing Vehicular Ad-Hoc Networks (VANETs). These networks leverage the communication capabilities of vehicles to enhance traffic management, improve road safety, and optimize overall transportation efficiency \mbox{\cite{Chhabra2021Krishna}}. Fig. \mbox{\ref{fig:SG_arch}} shows the Conceptual framework of a VANET architecture. VANET is a key ITS technology that can be leveraged to alleviate traffic congestion. VANET vehicles have onboard computing and communication modules to disseminate critical traffic information, navigation, and road services updates. Unlike traditional navigation applications that rely on user reports to provide real-time traffic conditions, often reacting to congestion after it has occurred, VANET-based traffic management systems excel by predicting future traffic flow. This proactive approach leverages reports from drivers about their current and intended destinations, offering a significant advantage in anticipating and mitigating potential traffic congestion issues. The predictive capability of VANETs distinguishes them from conventional navigation tools, which are not only reactive but also prone to bias in user-reported traffic conditions. Thus, contemporary research focuses on designing preventive techniques that leverage VANET to reduce congestion\mbox{\cite{rabieh2016privacy, zhang2019privacy, zhang2017efficient, lee2021vanet}}. These strategies require each motorist to report his or her current and future locations (i.e., travel routes) to a traffic management center or other adjacent vehicles.  This real-time collected information can then be aggregated and analyzed for traffic patterns, to generate a dynamic traffic density map (i.e., heat map) pinpointing likely congestion hotspots. Subsequently, the Traffic Management Center (TMC) recommend alternative routes to drivers via their VANET modules, alleviating congestion and improving traffic flow.

Despite the need to obtain vehicle location data to forecast future traffic congestion, it is vital to recognize the high sensitivity of this data. The temporal route information of each driver qualifies as a behavioral biometric marker that is as unique as a fingerprint. It is rare to discover two drivers with identical temporal route patterns\mbox{\cite{zhang2019privacy}}. Knowing this temporal route information raises privacy concerns since motorists' identities, movement patterns, and other sensitive information could be disclosed. Noteworthy is the possibility of third parties, such as insurance companies and travel brokers, gaining access to this sensitive data for their advantage. Moreover, criminals may also utilize this information to determine the optimum times to commit crimes. Thus, it is of utmost importance to develop privacy-preserving route reporting schemes that ensure the confidentiality of the route data of motorists.

Various trials have been explored to minimize urban congestion, such as enhancing transportation infrastructure, charging traffic fines, offering route information, enforcing traffic regulations, and boosting public transportation \mbox{\cite{ma2020ridesharing, sun2020managing}}. However, addressing urban congestion while preserving the drivers' privacy remains a complex and ongoing challenge. Despite the efforts mentioned above, several factors contribute to the persistence of congestion in urban areas. These include rapid population growth, increased vehicle ownership, and the allure of urban centers for economic opportunities. Short-term traffic flow forecasting primarily focuses on predicting the traffic flow condition in a few or hundreds of minutes. It is currently one of the most prominent research areas in ITS. Nevertheless, due to the stochastic and dynamic nature of traffic, traditional short-term traffic flow forecasting methods often face limitations in accuracy and reliability. For instance, because of the volatility and fluctuations caused by weather and the environment, determining the complicated nonlinear relationship between traffic flow data and time is exceedingly challenging.
Nonetheless, with the advent of Deep Learning (DL) algorithms for traffic flow prediction, transportation research has lately seen a resurgence in interest. \mbox{\cite{huang2014deep, lv2014traffic, ma2017learning, du2019deep, zheng2020hybrid}}. Existing works for estimating traffic flow employing DL possess certain shortcomings. Some studies employ a rudimentary neural network model, like Stacked Autoencoder (SAE), Long Short Term Memory (LSTM), or Convolutional Neural Network (CNN), which can only capture a fraction of the complicated characteristics of traffic flow. Although several works propose hybrid DL algorithms that combine distinct models to capture different features for traffic flow prediction, including spatial, temporal, and periodic features, they are frequently handled independently. Additionally, these efforts do not fully utilize the intricate structures present in traffic flow data. Furthermore, despite their potential benefits in enhancing traffic forecasting and management, these endeavors have often overlooked drivers' privacy.

To address these concerns, we propose an efficient and effective privacy-preserving location reporting and traffic forecasting scheme that protects the sensitive location information of drivers. Our proposed scheme divides the traffic management area into cells (geographic regions), each assigned a unique identification number (ID). Drivers must report their encrypted location information.
The encrypted information is subsequently aggregated by the TMC and employed as input for a multilayer DL model in traffic flow prediction. This model identifies and extracts hidden characteristics within traffic flow data,  constructing a traffic density map (i.e., heat map) that highlights probable regions of congestion. Drivers can then reroute their journey to avoid these regions. It is worth noting that the encrypted reports and decryption are performed utilizing a variety of functional encryption and decryption keys. Moreover, employing aggregation protects the temporal location information of each driver while revealing the bare minimum of information required to generate a traffic density map. The proposed research bridges a significant gap in designing an efficient VANET traffic management system, particularly in major and mid-sized cities experiencing rapid urbanization and traffic congestion. The primary contributions of this work are enumerated as follows:

\begin{enumerate}
    \item We propose a novel privacy-preserving location reporting scheme for traffic management systems, based on Inner Product Functional Encryption (IPFE)\mbox{\cite{boneh2011functional}}. This scheme incorporates advanced functional encryption and decryption techniques to safeguard the privacy of driver route information, while allowing access to specific encrypted data. Consequently, this approach maintains the confidentiality of sensitive data, while facilitating the prediction of future traffic congestion and enabling the creation of accurate traffic forecast density maps. This ensures that privacy is upheld without compromising the effectiveness of traffic management strategies.
    \item We developed a DL-based model for predicting traffic flow, integrating a hybrid architecture that combines Conv-LSTM and Bi-LSTM models with a Squeeze-and-Excitation (SE) module. The Conv-LSTM component captures spatial and short-term temporal patterns, while the Bi-LSTM is engineered to extract long-term temporal features, including daily and weekly cycles, reflecting the dynamic trends in traffic flow. Additionally, the inclusion of the Squeeze-and-Excitation (SE) module enhances the forecast performance by improving feature representation and discrimination.
    \item Our proposed scheme underwent comprehensive evaluation using both synthetic and real-world traffic data. The evaluation was divided in two parts: the first assessed the efficacy and overhead associated with the privacy-preserving route reporting scheme. The second part focused on measuring the accuracy of the traffic forecast model, with a comparative analysis of our results against both historical and contemporary research findings.
\end{enumerate}

This paper is organized as follows in the remaining sections. We describe in detail all related work in Section II. Then, in Section III, we outline the system models and the design objectives. Section IV details several DL models, techniques, and methods employed in our proposed scheme. Section V presents a thorough overview of our proposed privacy-preserving traffic forecast system. The privacy and security analysis and performance evaluation are provided in sections VI and VII, respectively. Finally, Section VIII summarizes the conclusions drawn from our study. 
\section{Related Work} \label{sec:Related Work}
% \hl{I worked on the entire Related Work section, used existing literature to find gaps in the topic and also show why IPFE is better qualitatively}
\subsection{Privacy-Preserving Route Reporting}
Recent advancements in literature have introduced a variety of privacy-preserving route reporting mechanisms for ITS \mbox{\cite{rabieh2015privacy, BBFog_Based, EBCPA_9749919, PP_CA9928430, LSM_TC10324334}}. Most of these studies utilize pseudonyms, homomorphic encryption, and k-anonymous algorithms,  further enhanced by blockchain technologies to bolster data integrity and privacy. These innovations, detailed in studies\mbox{\cite{calandriello2007efficient, rabieh2016privacy, BBFog_Based, EBCPA_9749919}}, aim to enhance data integrity and privacy. Furthermore, other traffic management techniques leverage transferable Federated Learning (FL) and Graph Convolutional Network (GCN) approaches\mbox{\cite{LSM_TC10324334, xia2022short}} for crowdsensed data,  have emerged as cutting-edge solutions for addressing the challenges of data scarcity and improving traffic management efficiency while safeguarding the privacy of crowdsourced data in ITS. While these strategies significantly contribute to preserving user identity and sensitive information protection, they encounter notable limitations including scalability issues, system complexity, blockchain overhead, heightened security vulnerabilities, and dependency on internet connectivity. Moreover, these strategies may incur substantial costs and present considerable barriers during their adoption and integration within the existing traffic management infrastructure. As such, while promising, these schemes may fall short in addressing the nuanced demands of dynamic, real-time traffic management scenarios, underscoring the need for continued innovation and adaptation in this rapidly evolving field.

\subsection{Deep Learning for Traffic Forecasting}
Traffic flow forecasting initially included three primary model types: parametric, non-parametric, and hybrid. Parametric models like ARIMA excel in analyzing time series data for traffic forecasting on expressways and urban roads\mbox{\cite{levin1980forecasting, hamed1995short}}, with innovations such as Kohonen-ARIMA (KARIMA)\mbox{\cite{van1996combining}} subset ARIMA\mbox{\cite{lee1999application}}, and seasonal ARIMA\mbox{\cite{williams2003modeling}} enhancing their precision for nonlinear data. Non-parametric models, including K-Nearest Neighbor (KNN) and Support Vector Regression (SVR)\mbox{\cite{feng2018adaptive}}, adapt well to complex data relationships but can face optimization hurdles and susceptibility to local minima.. Hybrid models combine the strengths of both, using techniques from ARIMA, Empirical Mode Decomposition (EMD), Singular Value Decomposition (SVD), and Neural Networks (NNs) to achieve superior accuracy and robustness in predicting traffic flow\mbox{\cite{10145923, tan2009aggregation}}.
DL further advances traffic flow prediction with Lv et al.\mbox{\cite{lv2014traffic}} showcasing the effectiveness of Stacked Autoencoders (SAEs) in surpassing traditional methods like Support Vector Machines (SVMs) and Feedforward Neural Networks (FNNs) in estimating traffic flows. DL models in traffic flow forecasting, can be segmented into short-term, long-term, and hybrid models. For short-term predictions, DL models incorporating CNNs, GCNs, and their variants have been effective in capturing spatial-temporal traffic patterns\mbox{\cite{SPTGCN9945663, Inception10032279, CNNGRU9701439, ma2022novel, ma2017learning}}, yet they struggle with temporal sequence data, where past information crucially predicts future outcomes. The introduction of LSTM networks by Tian et al.\mbox{\cite{tian2015predicting}} highlighted their superiority in capturing temporal dynamics, paving the way for subsequent variants\mbox{\cite{zhaowei2020short, ma2021short}} that further illustrate LSTMs' proficiency in long-term forecasting. However, these models often overlook the impact of road network layouts. Hybrid models\mbox{\cite{cheng9345387, zheng2020hybrid}} merging CNNs for spatial insight and LSTMs for temporal analysis have markedly improved traffic prediction, merging the strengths of both to enhance traffic management. Nonetheless, the success of these sophisticated models hinges on the quality and availability of traffic data. The process of data collection and analysis, especially from motorists and connected vehicles, raises significant user privacy and data security concerns, necessitating stringent data protection protocols that adds complexity to these forecasting systems.

Despite the scarcity or limited endeavors in both research and development to fully address the dual challenges of ensuring user privacy and data security in traffic data collection and creating dependable traffic forecasting systems, existing studies offer promising directions. For instance, Xia et al. \mbox{\cite{xia2022short}} present a system that combines GCN with FL for modeling traffic patterns. This approach utilizes GCN for identifying spatial dependencies in traffic data and employs FL for privacy-preserving collaborative learning without sharing raw data. Though ingenuous, this innovative system grapples with hurdles, including scalability issues, communication bottlenecks, susceptibility to adversarial threats, integration complexities with existing systems, and limited adaptability across different environments, highlighting the need for further research. To overcome the limitations identified in existing research and offer a holistic solution, we introduce a novel, lightweight privacy-preserving traffic forecasting system. Our system uniquely leverages functional encryption based on cryptography for scalable, internet-independent, and efficient privacy-preserving solution. It enables intricate encryption and computation on encrypted traffic data, safeguarding data security and privacy without compromise. Further enriching our solution, we incorporate a hybrid Conv-LSTM and Bi-LSTM model with an SE module, enhancing the extraction and analysis of crucial temporal-spatial dynamics, alongside short-term and long-term traffic patterns. This approach significantly boosts the forecast accuracy and precision, setting a new benchmark for traffic forecasting systems in terms of privacy preservation and operational efficiency with exceptional forecast reliability.

%%%%%%%%%%%%%%%%%%%%%%%%%%%%%%%%%%%%%%%%%%%%%%%%%%%%%%%%

%%%%%%%%%%%%%%%%%%%%%%%%%%%%%%%%%%%%%%%%%%%%%%%%%%%%%%%%

% \section{System Models}

\section{System Models and Design Objectives}  \label{sec:System Models}

This section provides an overview of the system model, which includes the network model, threat model, and the proposed scheme's design goals.
%%%%%%%%%%%%%%%%%%%%%%%%%%%%%%%%%%%%%%%%%%%%%%%%%%%%%%%%

\subsection{Network Model} \label{subsec:Network Model}  As shown in Fig.~\ref{fig:network_model}, our considered network model includes three main entities: the vehicle-side (drivers), traffic management center (TMC), and a key distribution center (KDC). The role of each entity is described below.
\begin{itemize}
    \item \textit{Drivers (D)}: As primary components of the traffic management system, each vehicle $D$ sends its encrypted location information periodically to the TMC. Communication between drivers and the TMC is either direct or indirect through a gateway (Roadside unit). A set of $D$, $\mathbb{D}=\{D_i, 1 \leq i \leq |\mathbb{D}|\}$, form the network.

    \item \textit{TMC}: As the central control and monitoring hub, the TMC uses encrypted location information from drivers for traffic flow analysis, congestion detection, and route planning in real time.

    \item \textit{KDC}: The KDC is a crucial offline entity responsible for preserving secure communication and data privacy by providing drivers $D$ and the TMC, respectively, with unique encryption and functional decryption keys.
    %In practice, KDC can be operated by a government agency like the Department of Motor Vehicles (DMV).

\end{itemize}

\vspace*{-3mm}
\subsection{Threat Model}\label{subsec:Threat Model}

We assume an honest but curious model for our system in which the TMC is honest in computing the true traffic conditions. However, it may attempt to learn sensitive information (i.e., driver identity and movement patterns) from the driver data it receives to its advantage. We also assume that the TMC is managed and operated by the Department of Motor Vehicles. We also assume that all the traffic data sent by the drivers is anonymized so that no personally identifiable information is revealed. The anonymization scheme used by the vehicles is out of the scope of this paper. Our system aims to implement a privacy policy that clearly outlines how data is collected, used, and shared. A driver (or group of drivers) may also be interested in deducing sensitive information about other drivers for their benefit. Furthermore, external adversaries $\mathcal{A}$ can operate individually or in collusion to launch attacks on the communication gateway between drivers and the TMC to gather sensitive information for their gains. Though the TMC and its communication with drivers are not fully trusted, the TMC maintains a high level of security and avoids collusion efforts between it and drivers because it is supervised and monitored by government organizations.

\subsection{Design Goals}\label{subsec:Design Goals} In our proposed scheme, we anticipate achieving the following objectives.
\begin{itemize}
    \item \textit{Privacy Preservation}: Our design seeks to develop robust mechanisms that protect the location and identity of drivers (via unauthorized access and monitoring prevention) while enabling effective traffic management and congestion mitigation.

    \item \textit{Real-time Traffic Forecasting}: By leveraging advanced predictive models' accuracy and real-time traffic forecasting capabilities, our system aims to provide reliable and informed traffic data, enabling proactive congestion management and efficient route planning.

    \item \textit{Scalability and Efficiency}: Our system is designed to be scalable for real-time deployment and operation under dynamic traffic conditions. This scalability extends to accommodating increasing drivers, expanding map sizes, and handling various network loads, ensuring efficient performance with minimal latency and computational overhead.

    \item \textit{Secure Communication}: With secure communication channels, our scheme utilizes secure protocols and cryptography techniques to guarantee the integrity, security, and confidentiality of data exchanged between system entities.

\end{itemize}

\section{Preliminaries} \label{sec:Preliminaries}
\subsection{Functional Encryption}
\textit{Functional encryption} (FE) refers to a type of cryptography that allows for the encryption of a message $\boldsymbol{x}$ using a key $k$ to get $Enc_k(\boldsymbol{x})$, as well as the ability of a designated decryptor to compute the output of a function $f$ on the encrypted message using a decryption key $dk$ without being able to learn the message itself (i.e., $Dec_{dk}(Enc_k(x))=f(x)$)~\cite{10.1007/978-3-642-19571-6_16}.

Recently, the focus on FE has been increasing, especially on how to design efficient schemes for limited classes of functions or polynomials, such as linear~\cite{10.1007/978-3-662-53015-3_12,10.1007/978-3-662-46447-2_33} or quadratic~\cite{10.1007/978-3-319-63688-7_3}. In this paper, we focus on a specific type of functional encryption known as inner product functional encryption (IPFE)\mbox{\cite{boneh2011functional}}, which allows for the computation of the inner product of two encrypted vectors.
%, and it has been built under well-understood security assumptions~\cite{10.1007/978-3-662-46447-2_33}.
%In an IPFE scheme, given the encryption of a vector $\boldsymbol{x}$ and a functional decryption key associated with a vector $\boldsymbol{y}$, one can obtain only the dot product result $(\boldsymbol{x} \cdot \boldsymbol{y})$ by decrypting the encryption of $\boldsymbol{x}$ and without being able to learn $\boldsymbol{x}$. IPFE consists of three parties as follows.
In an IPFE framework, when provided with the encryption of a vector $\boldsymbol{x}$ and a functional decryption key linked to a vector $\boldsymbol{y}$, one can exclusively derive the dot product result $(\boldsymbol{x} \cdot \boldsymbol{y})$ by decrypting the encrypted form of $\boldsymbol{x}$—all without gaining access to the actual values of $\boldsymbol{x}$. IPFE involves three distinct parties, outlined as follows.
\begin{itemize}
    \item \textit{KDC}: The KDC produces an encryption key for the encryptor and a functional decryption key for the decryptor.

    \item \textit{Encryptor}: The encryptor encrypts the plaintext vector $\boldsymbol{x}$ into the ciphertext and sends it to the decryptor.

    \item \textit{Decryptor}: The decryptor uses the functional decryption key $dk_y$ obtained from the KDC to evaluate and access $(\boldsymbol{x} \cdot \boldsymbol{y})$, where  $\boldsymbol{x}$ and $\boldsymbol{y}$ are the plaintext vector and the encrypted vector, respectively. The decryptor is obliged to maintain non-collusion with the KDC.
\end{itemize}
\subsection {Convolution/ LSTM and Bi-LSTM}
CNN and LSTM are powerful deep-learning architectures widely used in computer vision and natural language processing. CNNs use a combination of convolutional layers, pooling layers, and fully-connected layers to extract features from an image and then classify the image into one of the predefined classes. CNNs are particularly suitable for object recognition, facial recognition, and image segmentation tasks. On the other hand, LSTM networks are mainly used for natural language processing tasks such as language translation, sentiment analysis, and text generation. LSTM networks are composed of multiple layers of memory units, which are responsible for storing information from the past and using it to make predictions. They are particularly powerful when understanding data sequences, such as sentences, and predicting what comes next. A combination of CNN and LSTM, known as Conv-LSTM, is usually used to improve the performance of a neural network. The wide adoption of Conv-LSTM is due to their high accuracy. The purpose of using attention-based Conv-LSTM is to make the near-future predictions accurate and timely.
\begin{figure*}[t]
    \centering
    \includegraphics[width=1.0\textwidth]{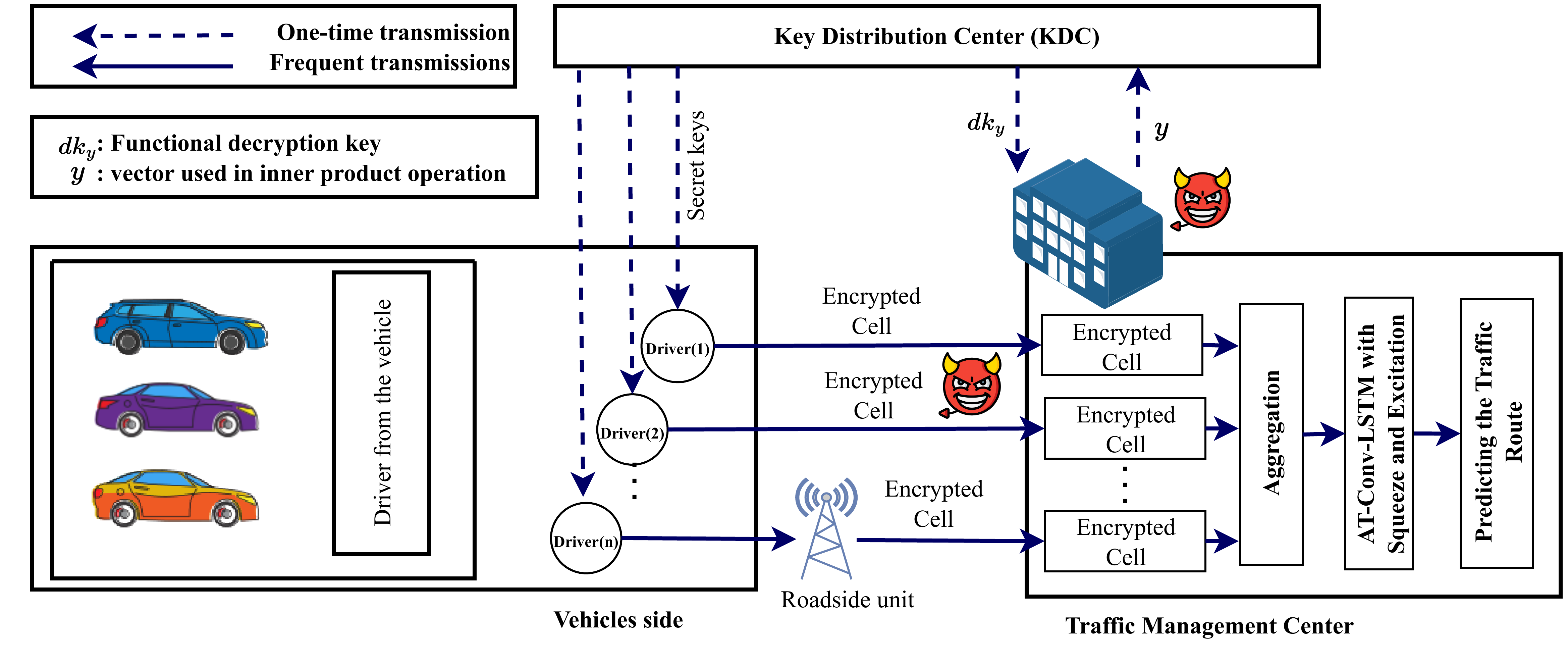}
    \caption{Illustration of the Privacy-Preserving Traffic Management System}
    \label{fig:network_model}
\end{figure*}
\subsection{Attention Mechanism}
An attention mechanism allows deep learning models to selectively focus on certain parts of the input when making predictions. It is particularly useful in natural language processing and image recognition tasks. In these tasks, the model must be able to identify and understand specific parts of the input to make accurate predictions. The attention mechanism is implemented by adding an attention layer to the neural network, which learns to assign weights to different input parts. These weights are then used to create a weighted sum of the input, which is then passed to the next network layer. Attention mechanisms have been shown to improve the performance of neural networks on a wide range of tasks and are now widely used in many state-of-the-art models. An attention-based Conv-LSTM combines attention mechanisms and Conv-LSTMs to provide accurate forecasting.
\subsection{Squeeze-and-excitation}
Squeeze-and-excitation (SE) \cite{hu2018squeeze} is a type of attention mechanism that aims to improve the feature representation of a neural network. It works by first compressing the feature maps' spatial dimensions, reducing the number of channels. The resulting feature maps are then passed through an excitation module, which learns to assign weights to different channels based on their importance. These weights are then used to recalibrate the feature maps, improving the network's overall feature representation. SE has been shown to improve the performance of neural networks on various tasks such as image classification, object detection, and semantic segmentation, particularly in architectures like CNNs. It can be added to existing architectures like CNNs or convolutional LSTMs as a module.

%%%%%%%%%%%%%%%%%%%%%%%%%%%%%%%%%%%%%%%%%%%%%%%%%%%%%%%%

\section{Proposed Scheme} \label{sec:Proposed Scheme}
As depicted in Figure \ref{fig:network_model}, our proposed framework comprises two primary components: 1) Privacy-Preserving Location Reporting and Aggregation for Drivers, and 2) Traffic Forecasting through Deep Learning. The first component encompasses system initialization, driver location reporting, and server-side aggregation of information for traffic monitoring. The second component involves a deep learning-based traffic forecasting algorithm. Our model utilizes Conv-LSTM on aggregated driver data to predict short- and long-term traffic patterns while ensuring driver privacy. Additionally, our model incorporates an attention mechanism and a squeeze-and-excitation block, significantly improving performance. The following subsections explain the details of each building block.

\subsection{Drivers Location Reporting and Aggregation}
\begin{figure}[!ht]
        \captionsetup{labelfont=it,justification=centering}
        \centering
           \includegraphics[scale=.10]{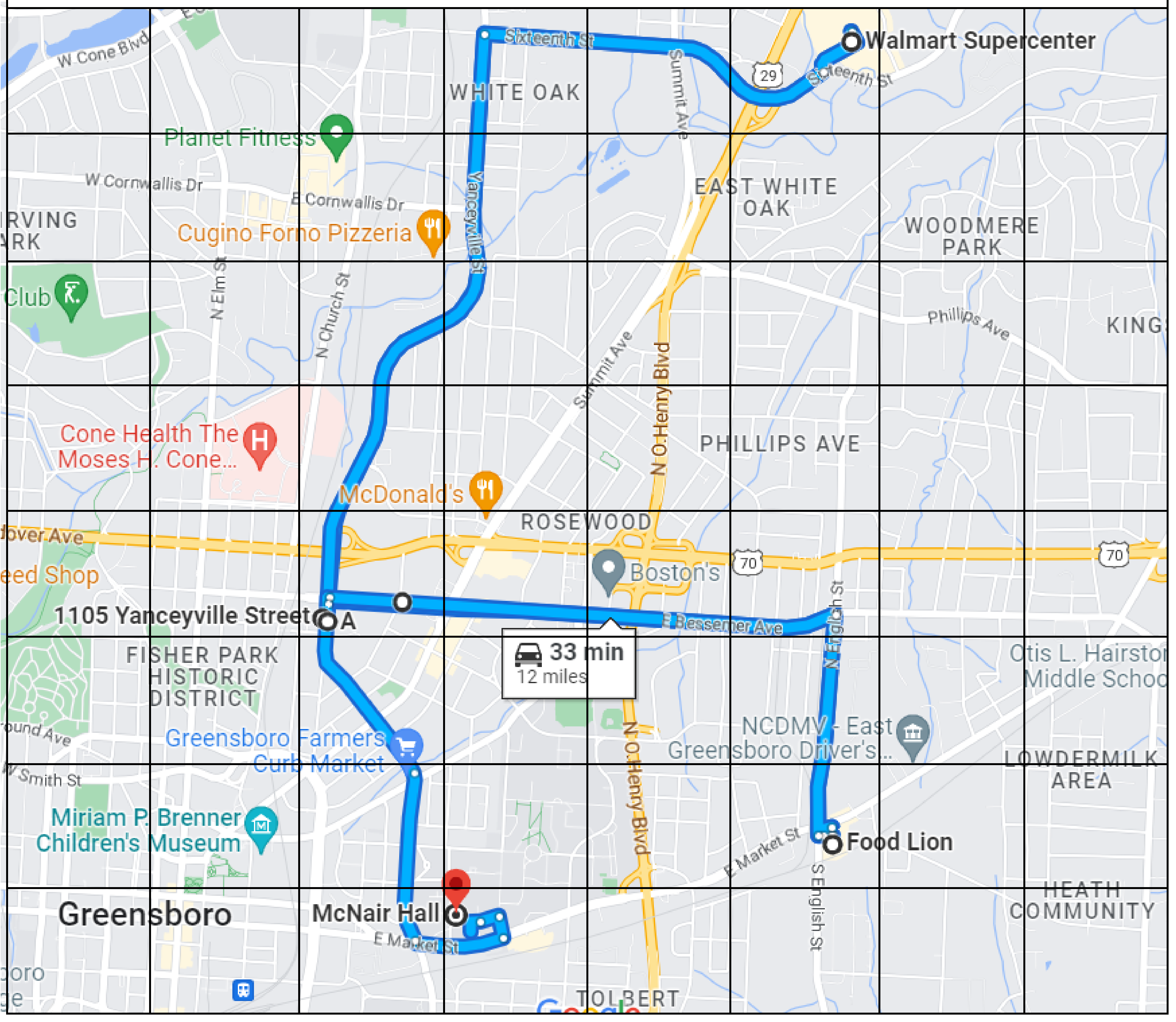}
           \caption{A traffic management area partitioned into distinct geographic zones (i.e., cells).}
           \label{fig:Cell map}
\end{figure}

We assume that the traffic management area is divided into a set of geographic areas called cells, as illustrated in Figure \ref{fig:Cell map}. Each cell is assigned a unique identifier, similar to zip codes.

%%%%%%%%%%%%%%%%%%%%%%%%%%%%%%%%%%%%%%%%%%%%%%%%%%%%%%%%

%%%%%%%%%%%%%%%%%%%%%%%%%%%%%%%%%%%%%%%%%%%%%%%%%%%%%%%%

\begin{itemize}
    \item To report their location, each $D_i \in \mathbb{D}$, where $\{1 \leq i \leq |\mathbb{D}|\}$,  employ IPFE scheme\mbox{\cite{boneh2011functional}} to conceal their association with a specific cell, denoted as $\boldsymbol{l}^j_i[t] = 1$, where ${1 \leq j \leq |\mathcal{L}|}$, and $\mathcal{L}$ represents the total number of grid cells within a given reporting area. Additionally, drivers encrypt the remaining ($k-1$) dummy cells with a value of zero to maintain k-anonymity~\cite{sweeney2002k}. The outcome is a set of $k$ ciphertexts, labeled as $C^1_i[t]$ through $C^k_i[t]$, which are subsequently transmitted to the decryptor (i.e., TMC). This encryption mechanism safeguards the confidentiality of the driver's precise location.
    
    \item At every time slot $t$, the TMC receives encrypted cell information $C^j_i[t]$ from all drivers and uses the functional decryption key $\boldsymbol{dk}$ to obtain the aggregated encrypted drivers' density for each cell $j$ (i.e., $Dec\boldsymbol{_{dk}}([C^j_1[t], \dots, C^j_{|\mathbb{D}|}[t]])=\sum_{i=1}^{|\mathbb{D}|} \boldsymbol{l}^j_i[t],$ where $\boldsymbol{l}^j_i[t]$ is the plaintext status of the grid cell $j$ sent by $D_i$ at time slot $t$). If the TMC receives fewer ciphertexts for cell $j$ than the total number of drivers $|\mathbb{D}|$,  the TMC adds dummy ciphertexts containing encrypted zeros to compensate for any missing reports from drivers for a particular cell.
\end{itemize}

The main phases of the route report are described as follows.

\begin{table}[!t]
	\centering
	\caption{Main notations.}
	\label{tab:multi_notation}
	
	\begin{tabular}{cl}
		
		\hline
       Notation        &     Description\\
       \hline
       $\mathbb{D}$    &    Number of drivers\\
       $\mathbf{p k}_i$    &    Encryption keys of Drivers\\
       $\mathcal{L}$    &    Total number of grid cells\\
       $l^j_i[t]$    &      Status of cell $j$ reported by driver $D_i$ at time  $t$\\
       $C^j_i[t]$    &    Encrypted status of cell $j$ reported by $D_i$ at time $t$\\
       $\mathbb{G}, p, g$  & Public parameters for the functional encryption\\
       $\boldsymbol{dk}$ & Functional decryption keys\\
       $\textbf{X}[t^s]$ & Current traffic density over $t^{s}-n,..., t^{s}$ \\
       $\textbf{X}[t^d]$ & Daily historical traffic density over $t^{d}-n,..., t^{d}$ \\
       $\textbf{X}[t^w]$ & Weekly historical traffic density over $t^{w}-n,..., t^{w}$ \\
       $G[t^{s}]$ & Output from the CNN\\
       $H_{2}[t^{s}]$ & The LSTM hidden state indicating the spatial-temporal\\ & feature for time step $t^s$.\\
       $C, H$    &    Channel and spatial dimensions of the Squeeze operation\\
       $G[t^s], G^{'}[t^s]$    &    Output of CNN and Squeeze and excitation\\
       $H_a[t^{s}]$ & The output of Conv-SE-LSTM at each time step $t^s$\\
       $\tau$ & Time interval\\
       $\beta_{k}$ & The attention value\\

   % Please do not add more symbols to the table until the corresponding section is finalized

   %    $\pi$ &    Proof of correct encryption\\
   %    $\kappa$ &    AES encryption key\\
       % $CT$ &    Encr\\
       \hline
	\end{tabular}
	% \vspace{-5mm}
\end{table}

\subsubsection{System Initialization}

During the system initialization, the KDC computes and distributes the following:
(a) Public parameters;
(b) Driver's encryption keys;
and (c) TMC's functional decryption keys.

%%%%%%%%%%%%%%%%%%%%%%%%%%%%%%%%%%%%%%%%%%%%%%%%%%%%%%%

\textit{{a) Public Parameters Generation:}}
To generate the public parameters, the KDC should:

$\operatorname{Setup}\left(1^\lambda, \mathcal{F}_\mathbb{D}\right)$ : The algorithm first generates secure parameters as $\mathcal{G}:=(\mathbb{G}, p, g) \leftarrow \operatorname{GroupGen}\left(1^\lambda\right)$, and then generates several samples as $a_i \leftarrow{ }_R \mathbb{Z}_p^1, \mathbf{a}_i:=(1, a_i)^{\top}, \forall i \in\{1, \ldots, |\mathbb{D}|\}$, in addition to $\mathbf{W}_i \leftarrow_R \mathbb{Z}_p^{1 \times 2}, u_i \leftarrow_R \mathbb{Z}_p^{1}$. Then, it generates the master public key and master private key as

$\operatorname{\mathbf{mpk}}:=(\mathcal{G},[\mathbf{a}_i]\footnote[1]{Note that $[x]=g^x$. In our representation, we adopt the bracket notation implicitly from \cite{escala2017algebraic}, which is widely recognized and used as a standard in the cryptographic community.},\mathbf{W}_i \mathbf{a}_i]), \mathbf{m s k}:=(\mathbf{W}_i, u_{i})_{i \in\{1, \ldots, |\mathbb{D}|\}}
 $
        % \end{itemize}
%%%%%%%%%%%%%%%%%%%%%%%%%%%%%%%%%%%%%%%%%%%%%%%%%%%%%%%
\vspace*{2mm} \\
\textit{b) Drivers' Encryption Keys Generation:}
KDC constructs and distribute $|\mathbb{D}|$ encryption keys to the drivers in the network as follows:  $\mathbf{p k}_i:=\left(\mathcal{G},\left[\mathbf{a}_i\right],\left[\mathbf{W}_i \mathbf{a}_i\right], u_i\right)$.
%$\boldsymbol s_{i} \in$ $\mathbb{Z}_{q}^{2}$, where $\boldsymbol s_{i}$ is the secret key of $D_i$, for $1 \leq i \leq |\mathbb{D}|$, and $|\mathbb{D}|$ denotes the number of drivers in the network.
%Then, the secret key is $\{\boldsymbol s_{i}\}_{\forall_{i}}$.
\vspace*{2mm} \\
\textit{c) TMC's Functional Decryption Key Generation:}
% $\boldsymbol{dk}$ is the functional decryption key for aggregating the drivers' route information.
%Each key is a set of decryption keys.
% The KDC generates these functional decryption key $\boldsymbol{dk}$.
% \renewcommand{\theenumi}{\roman{enumi}}
% \begin{enumerate}
%      \item
     The KDC uses a vector of ones, denoted as $\mathbf{y}^{1\times|\mathbb{D}|}$, which has a length equal to the number of drivers in the network, in order to compute the functional decryption key $\boldsymbol{dk}$. The purpose of using this vector is to ensure that when an inner product is performed with the driver's report, all drivers for grid cell $j$ and aggregated reports for cell $j$ are obtained. The generation process for the $\boldsymbol{dk}$ key is outlined below.

     The KDC performs the following operation to compute the functional decryption key $\boldsymbol{dk}$ as:
     $$
\boldsymbol{dk}:=  \mathbf{d}_i^{\top} \leftarrow \left( y_i \mathbf{W}_i \right)_{i\in |\mathbb{D}|}  , z \leftarrow \sum_{i\in |\mathbb{D}|}  y_i {u}_i
         %  \boldsymbol{dk} = \sum_{i=1}^{|\mathbb{D}|} \boldsymbol s_{i}  \boldsymbol{y}[i]  = \sum_{i=1}^{|\mathbb{D}|} \boldsymbol s_{i} \in \mathbb{Z}_{q}^{2},
$$

    This equation is equivalent to aggregating the secret keys from all drivers to generate $\boldsymbol{dk}$. Then, the KDC sends the $\boldsymbol{dk}$ to the TMC.

% \end{enumerate}

%%%%%%%%%%%%%%%%%%%%%%%%%%%%%%%%%%%%%%%%%%%%%%%%%%%%%%%

%\subsection{Reporting Fine-grained Power Consumption Readings} \label{sub:Report Generation}
\subsubsection{Reporting Drivers Locations}

For each reporting period $t$, driver $D_i$ encrypts the cell $j$ information, $\forall 1<j<|\mathcal{L}|$, and generate the ciphertext $C^j_i[t]$. This encryption ensures that the cell information is kept private and only authorized parties can access it. Each cell information is encrypted separately, allowing the TMC to compute the aggregated reports for cell $j$ without learning the individual reports themselves. The encrypted cell information is generated as follows.

$\operatorname{Encrypt}\left(\mathbf{p k}_i, \boldsymbol{l}_i^{j}\right)$ : The algorithm first generates a random nonce ${r_i^{j} \leftarrow{ }_R \mathbb{Z}_p}_{j \in\{1, \ldots, K\}}$ and then computes the ciphertext as
$$
\mathbf{C}_i^{j}[t]:=(\left[{t}_i^j\right] \leftarrow[\mathbf{a}_i {r}_i^{j}], \left[\mathbf{c}_i^j\right] \leftarrow [\boldsymbol{l}_i^{j}[t]+u_i+ \mathbf{W}_i \mathbf{a}_i r_i^{j}]) .
$$

It should be noted that the drivers do not need to report the encryption status for all cells within the reporting area. Instead, they can employ K-anonymity~\cite{sweeney2002k} to selectively report only a subset of cells, thereby ensuring privacy and reducing computational overhead. This

%For each reporting period $t$, driver $D_i$ encrypts the cell $j$ information, $\forall 1<j<|\mathcal{L}|$, represented as $\ell^j_t \in \mathbb{Z}_p$, using its secret key $\boldsymbol{s}_i$ provided by the Key Distribution Center (KDC) to generate the ciphertext $C^j_i[t]$. This encryption ensures that the cell information is kept private and only authorized parties can access it. Each cell information is encrypted separately, allowing the TMC to compute the aggregated reports for cell $j$ without learning the individual reports themselves. The encrypted cell information is generated as follows.
%\begin{equation}\label{Enc_eq}
%        C^j_i[t]=(\boldsymbol{s}_{i}^{\top} \cdot \boldsymbol{U}_{t} )+\boldsymbol{l}^j_i[t]P \in \mathbb{G},
%\end{equation}
%Where: $\boldsymbol{U}_{t}=\mathcal{H}(t) \in \mathbb{G}^{2}.$

%%%%%%%%%%%%%%%%%%%%%%%%%%%%%%%%%%%%%%%%%%%%%%%%%%%%%%%

\subsubsection{Aggregating the Drivers Reports}After collecting all the ${D}$'s encrypted locations ($\mathbold{c}_{t}$) at reporting period $t$, where $\mathbold{c}_{t} = [C^j_1[t], C^j_2[t], \dots, C^j_{|\mathbb{D}|}[t] ]$, the TMC uses the functional decryption key $\boldsymbol{dk}$ to obtain the total aggregated location data for traffic by performing.

 %\begin{enumerate}
 %\item
 Given the functional decryption key $\boldsymbol{dk}$ and ciphertexts $\mathbold{c}_{t}$, the TMC can compute:
$$
=\frac{\prod_{i \in[\mathbb{D}]}\left(\left[\mathbf{y}^{\top} \mathbf{c}_i\right] /\left[\mathbf{d}_i^{\top} {t}_i\right]\right)}{[z]}
$$
$$
=\frac{\prod_{i \in[\mathbb{D}]}\left(\left[\mathbf{y}^{\top} \mathbf{c}_i\right] /\left[\mathbf{y}^{\top} \mathbf{W}_i \mathbf{a}_i r_i^{j}\right]\right)}{[z]}
$$
$$
=\frac{\prod_{i \in[\mathbb{D}]}\left(\left[\mathbf{y}^{\top}(\boldsymbol{l}_i^{j}[t]+\mathbf{u}_i+ \mathbf{W}_i \mathbf{a}_i r_i^{j})\right] /\left[\mathbf{y}^{\top} \mathbf{W}_i \mathbf{a}_i r_i^{j}\right]\right)}{[z]}
$$
$$
=\frac{\prod_{i \in[\mathbb{D}]} \left[\mathbf{y}^{\top}\boldsymbol{l}_i^{j}[t]+ \mathbf{y}^{\top} \mathbf{u}_i+\mathbf{y}^{\top} \mathbf{W}_i \mathbf{a}_i r_i^{j} - \mathbf{y}^{\top} \mathbf{W}_i \mathbf{a}_i r_i^{j}\right]}{[z]}
$$
$$
=\prod_{i \in[\mathbb{D}]}\left[\mathbf{y}^{\top}\boldsymbol{l}_i^{j}[t]+ \mathbf{y}^{\top} \mathbf{u}_i+\mathbf{y}^{\top} \mathbf{W}_i \mathbf{a}_i r_i^{j} - \mathbf{y}^{\top} \mathbf{W}_i \mathbf{a}_i r_i^{j}-\mathbf{y}^{\top} \mathbf{u}_i\right]
$$
$$
=\prod_{i \in[\mathbb{D}]}\mathbf{y}^{\top}\boldsymbol{l}_i^{j}[t]
$$
$$
=\sum_{i=1}^{|\mathbb{D}|} \boldsymbol{l}^j_i[t]
$$
 %$$

 %       \boldsymbol{U}_{t}=\mathcal{H}(t) \in \mathbb{G}^{2}.
 %$$

% \item Next, the TMC computes:
%\begin{equation}\label{decryption_monitoring}
%\begin{aligned}{} &\sum_{i=1}^{|\mathbb{D}|} C^j_i[t]  -\boldsymbol{U}_{t}^{\top}  \boldsymbol{dk}\\&=\sum_{i=1}^{|\mathbb{D}|} ((\boldsymbol{s}_{i}^{\top} \cdot \boldsymbol{U}_{t} )+\boldsymbol{l}^j_i[t] P) - (\sum_{i=1}^{|\mathbb{D}|} \boldsymbol s_{i})^{\top}  \boldsymbol{U}_{t}   \\ &=  (\sum_{i=1}^{|\mathbb{D}|} \boldsymbol s_{i})^{\top}  \boldsymbol{U}_{t} +\sum_{i=1}^{|\mathbb{D}|} \boldsymbol{l}^j_i[t]   P -(\sum_{i=1}^{|\mathbb{D}|} \boldsymbol s_{i})^{\top} \boldsymbol{U}_{t}  \\&= (\sum_{i=1}^{|\mathbb{D}|} \boldsymbol{l}^j_i[t]) P \in \mathbb{G}.  \end{aligned}
%\end{equation}
 % \item Finally, the TMC uses discrete logarithm to compute: $$\sum_{i=1}^{|\mathbb{D}|} \boldsymbol{l}^j_i[t] .$$
%\end{enumerate}
Solving the discrete logarithm is not a challenging task due to the relatively small value of  $(\sum_{i=1}^{|\mathbb{D}|} \boldsymbol{l}^j_i[t] )$. While many methods have been introduced to compute the discrete logarithm, such as Shank's baby-step giant-step algorithm~\cite{10.1007/3-540-69053-0_18}, we resorted to using a lookup table to compute it efficiently in a light-weight manner.

By performing the above steps, the result $(\sum_{i=1}^{|\mathbb{D}|} \boldsymbol{l}^j_i[t] )$ is
 the summation of the drivers passing through grid cell $j$ at each reporting period $t$. After the aggregation, the TMC can use the encrypted information to forecast traffic conditions, such as traffic density and congestion, as explained in the next section.

 \begin{figure*}[t]
        \centering
        \includegraphics[width=0.8\textwidth]{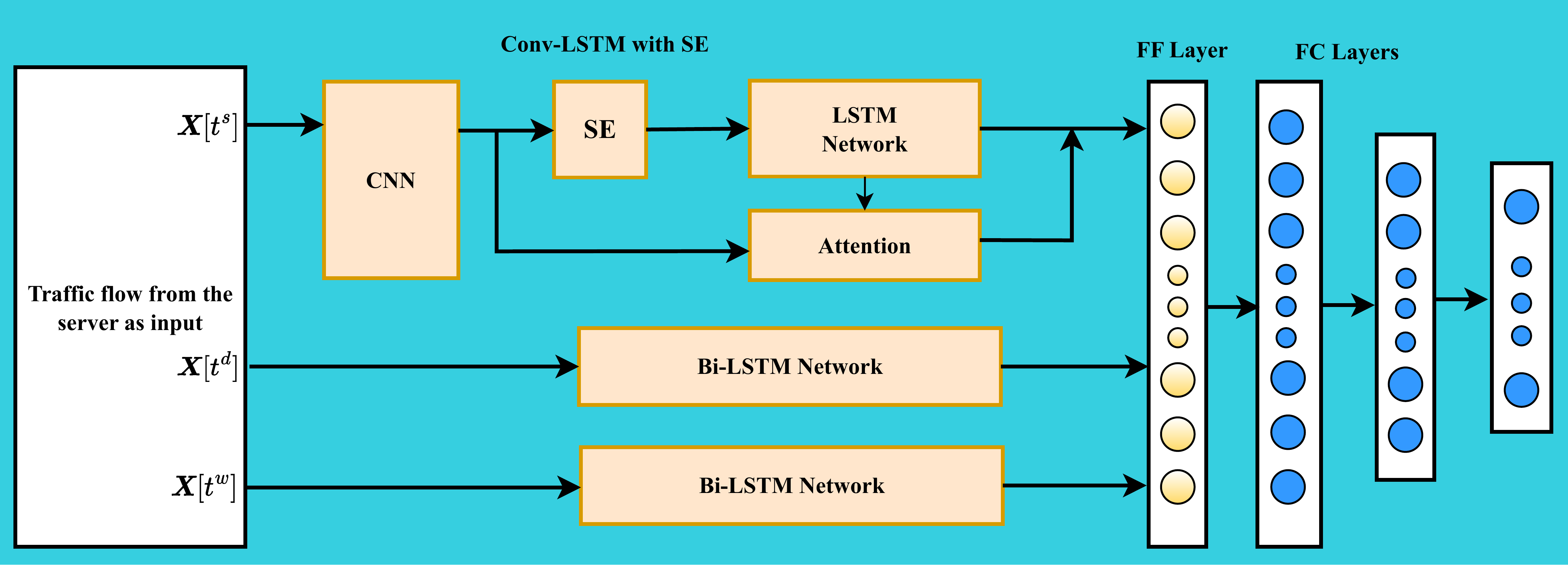}
        \caption{Architecture of Attention-Based Conv-LSTM Network.}  \label{fig:ffn_arch}
\end{figure*}

\subsection{Deep learning-based Traffic Forecasting}  
\begin{figure}[!ht]
        \centering
        \includegraphics[width=3.4in]{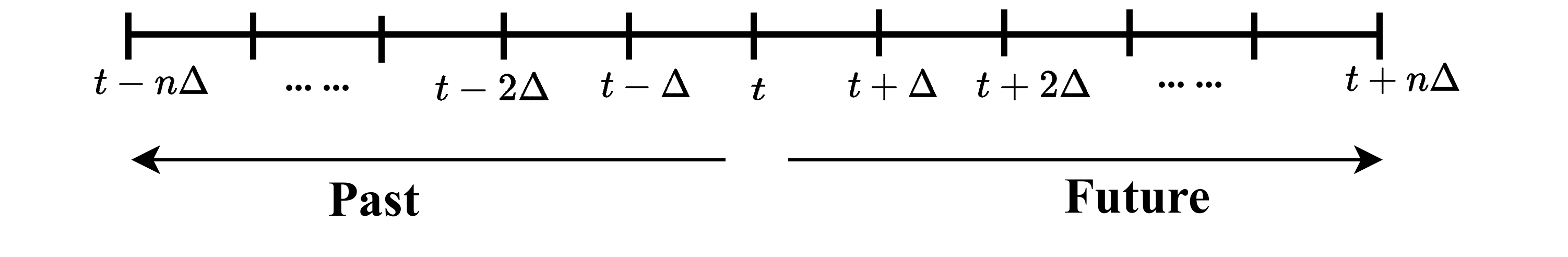}
        \caption{The Traffic forecasting time horizon.}
        \label{fig:time} \vspace{-5mm}
\end{figure}

\textbf{Traffic Flow Process Formulation:} The process of traffic flow prediction can be formulated mathematically as the drivers' density and congestion patterns within each cell under the traffic monitoring area. This formulation involves the analysis of historical density, real-time density, and future density. As shown in Figure \ref{fig:time}, at the current time $t$, the objective is to predict the traffic flow of a specific grid cell at the time interval $(t +  h \Delta)$ for a given prediction horizon, utilizing the past traffic status. Let $X^j[\tau]$ denote the traffic flow of the $j^{\text{th}}$ observation route during the $\tau^{\text{th}}$ time interval.  The traffic flow values $X^j[\tau]$ correspond to $\tau = t-n \Delta, \ldots, t-\Delta, t$.  Here, $\Delta = 5$ minutes, $n = 15$, and $h = 1, 3, 6, 12,$. This means that 75-minute historical data will be used to predict the traffic flow of the next 5, 15, 30, and 60 minutes.

% Let $X_{i}^{\tau}$ denote the traffic flow of the $i^{\text{th}}$ observation route during the $\tau^{\text{th}}$ time interval.  At a current time t, the task is to predict the traffic flow of a specific point route at a time interval $(t + \Delta h)$ for some prediction horizon.
% {$X_{i}^{\tau}$} $tau=t-n \Delta$,..., $t-\Delta$, $t$ and $i\in O$, where $O$ is the set of observation routes in the transportation network).

% \vspace*{-5mm}

% Here $ \Delta$ = 5 minutes, $n = 15$ and $h = 1, 3, 6, 12$, which means that 75-minute historical data will be used to predict the traffic flow of the next 5, 15, 30, and 60 minutes.
We create three spatiotemporal traffic flow matrices to capture the temporal and spatial aspects of traffic flow. This involves combining historical traffic flow data from neighboring locations at different time scales, including the current moment $t^{s}$, daily patterns $t^{d}$, and weekly trends $t^{w}$. The matrix $\boldsymbol{X}[t^s]$ specifically represents the current historical traffic density. It considers a time window spanning from $t^{s}-n$ to $t^{s}$ where each column of this matrix can be represented as the status of the reporting area at time $t^{s}$ denoted as $X[t^{s}]$ = $\left[\sum\limits_{i=1}^{|D|}l_{i}^{1}[t^{s}], \sum\limits_{i=1}^{|D|}l_{i}^{2}[t^{s}],...,\sum\limits_{i=1}^{|D|}l_{i}^\mathcal{L}[t^{s}] \right]^{T}$. The following matrix defines  $\boldsymbol{X}[t^s]$ with dimensions $\mathcal{L} \times n$, where $\mathcal{L}$ is the number of reporting cells, and $n$ is the size of the time window used for analysis.

% To construct the historical traffic flow, we define  $\sum\limits_{i=1}^{|D|}l_{i}^{j}[t]$ as the traffic flow at cell $j$ during time $t$.

% Consequently, the  traffic flow at time $t$ throughout the report area $\mathcal{L}$ can be represented as $X[t]$ = $\left[\sum\limits_{i=1}^{|D|}l_{i}^{1}[t], \sum\limits_{i=1}^{|D|}l_{i}^{2}[t],...,\sum\limits_{i=1}^{|D|}l_{i}^\mathcal{L}[t] \right]^{T}$. To form a spatial-temporal traffic flow matrix, we combine the historical traffic flows of the neighboring locations at the previous time instances. This involves constructing three matrices that represent the historical traffic densities at different time scales: current $t^{s}$, daily $t^{d}$, and weekly $t^{w}$. The following matrix $\boldsymbol{X}[t^s]$ represents the current historical traffic density over the time period $t^{s}-n, t^{s}-(n-1),..., t^{s}$. The dimension of this matrix is $\mathcal{L} \times n$.

% To construct the historical traffic flow, let $l_{t}^{l}$ note the traffic flow of an observation location l at time t. The historical traffic flow of the observation location l from time $t- n$ to $t$ can be represented as $X_{t}^{l}$ = $[l_{t-n}^{l}, l_{t-(n-1)}^{l},...,l_{t}^{l}]^{T}$.
% And ${\sum\limits_{j=1}^{|D|} l_{i}^{j}[t]}$ is the amount of cars passing through location $j$ at time $t$.
% Then, the combined historical traffic flow of its neighboring locations (total m locations including location l) forms a spatial-temporal traffic flow matrix as $\boldsymbol{X_{t}^{s}}$.

\begin{equation*}\label{Enc_eq}
{\begin{bmatrix} X[t^{s}-n] \\ . \\ . \\ . \\ .\\ X[t^s] \\ \end{bmatrix}}^T =\begin{bmatrix} {\sum\limits_{i=1}^{i=|D|} l_{i}^{1}[t^{s}-n]} & ..& {\sum\limits_{i=1}^{i=|D|} l_{j}^{1}[t^{s}]}\\ {\sum\limits_{i=1}^{i=|D|} l_{i}^{1}[t^{s}-n]} & ..& {\sum\limits_{i=1}^{i=|D|} l_{j}^{2}[t^{s}]} \\ .&.&..&\\ .&.&..&\\ .&.&..&\\ {\sum\limits_{i=1}^{i=|D|} l_{i}^\mathcal{L}[t^{s}-n]} &.. & {\sum\limits_{i=1}^{i=|D|} l_{j}^\mathcal{L}[t^{s}]}\\ \end{bmatrix}
  \end{equation*}

The next matrix defines the historical traffic densities with daily periodicity (i.e., in the previous day $d$) over the same time period $t^{d}-n,..., t^{d}, ..., t^{d}+n$.
The traffic data with daily periodicity can be obtained by considering the previous and following $n$ time intervals of the same moment as time $t^s$ from the preceding day. This can be represented as the matrix $\boldsymbol{X}[t^{d}]$.
      %\begin{equation}\label{Enc_eq}
%\boldsymbol{X_{t}^{d}}=\begin{bmatrix} f_{t^{d}-n}^{1} & %f_{t^{d}-(n-1)}^{1} & ... & f_{t^{d}}^{1}&...&f_{t^{d}+n}^{1}\\ %f_{t^{d}-n}^{2} & f_{t^{d}-(n-1)}^{2} & ... & %f_{t^{d}}^{2}&...&f_{t^{d}+n}^{2} \\ .&.&...&.&...&.  \\ %.&.&...&.&...&.  \\ .&.&...&.&...&. \\ f_{t^{d}-n}^{m} & %f_{t^{d}-(n-1)}^{m} & ... & f_{t^{d}}^{m}&...&f_{t^{d}+n}^{m} \\ %\end{bmatrix},
%  \end{equation}

\begin{equation*}\label{Enc_eq}
        {\begin{bmatrix} X[t^{d}-n] \\ . \\ . \\ . \\ .\\ X[t^d + n] \\ \end{bmatrix}}^T =\begin{bmatrix} {\sum\limits_{i=1}^{i=|D|} l_{i}^{1}[t^{d}-n]} & ..& {\sum\limits_{i=1}^{i=|D|} l_{j}^{1}[t^{d} + n]}\\ {\sum\limits_{i=1}^{i=|D|} l_{i}^{1}[t^{d}-n]} & ..& {\sum\limits_{i=1}^{i=|D|} l_{j}^{2}[t^{d} + n]} \\ .&.&..&\\ .&.&..&\\ .&.&..&\\ {\sum\limits_{i=1}^{i=|D|} l_{i}^\mathcal{L}[t^{d} -n]} &.. & {\sum\limits_{i=1}^{i=|D|} l_{j}^\mathcal{L}[t^{d}+ n]}\\ \end{bmatrix}
\end{equation*}

Similarly, the next matrix defines the historical traffic densities with weekly periodicity (i.e., in the previous week $'w'$) over the same time period $t^{w}-n,..., t^{w}, ..., t^{w}+n$.
Historical traffic flow data is constructed with weekly periodicity by considering previous and subsequent $n$ time intervals of the same moment as time $t^s$ in the last week as follows $\boldsymbol{X}[t^{w}]$.
 %     \begin{equation}\label{Enc_eq}
%\boldsymbol{X_{t}^{w}}=\begin{bmatrix} f_{t^{w}-n}^{1} & f_{t^{w}-(n-1)}^{1} & ... & f_{t^{w}}^{1}&...&f_{t^{w}+n}^{1}\\ f_{t^{w}-n}^{2} & f_{t^{w}-(n-1)}^{2} & ... & f_{t^{w}}^{2}&...&f_{t^{w}+n}^{2} \\ .&.&...&.&...&. \\ .&.&...&.&...&. \\ .&.&...&.&...&. \\ f_{t^{w}-n}^{m} & f_{t^{w}-(n-1)}^{m} & ... & f_{t^{w}}^{m}&...&f_{t^{w}+n}^{m} \\ \end{bmatrix},
%  \end{equation}

\begin{equation*}\label{Enc_eq}
{\begin{bmatrix} X[t^w-n] \\ . \\ . \\ . \\ .\\X[t^w+n] \\ \end{bmatrix}}^T =\begin{bmatrix} {\sum\limits_{i=1}^{i=|D|} l_{i}^{1}[t^{w}-n]} &  ... & {\sum\limits_{j=L}^{i=|D|} l_{j}^{1}[t^{w}+n]}\\ {\sum\limits_{i=1}^{i=|D|} l_{i}^{1}[t^{w}-n]} &  ... & {\sum\limits_{j=L}^{i=|D|} l_{j}^{2}[t^{w}+n]} \\ .&.&...& \\ .&.&...& \\ .&.&...&\\ {\sum\limits_{i=1}^{i=|D|} l_{i}^{L}[t^{w}-n]} &  ... & {\sum\limits_{j=L}^{i=|D|} l_{j}^{L}[t^{w}+n]} \\ \end{bmatrix},
\end{equation*}

% where $t^{w}$ denotes the same moment as time t in the last week.

\textbf{Deep Learning-based Forecasting:} The traffic forecasting model utilized by TMC is based on an attention-based Convolutional Sqeeze and Excitation and Long Short-Term Memory (Conv-SE-LSTM) deep learning architecture. The model's structure is depicted in Fig.~\ref{fig:ffn_arch}. The Conv-SE-LSTM module serves as the primary component of the proposed model, focusing on capturing the spatial-temporal features of traffic flow. The Conv-SE-LSTM module combines a CNN, a SE, and an LSTM network, as illustrated in Fig.~\ref{fig:ffn_arch}. The CNN component comprises two convolutional layers, while the LSTM component comprises two LSTM layers. The input to the Conv-LSTM module is a spatial-temporal traffic flow matrix denoted as $X[t^{s}]$, which represents the current historical traffic flow of the reporting area to be predicted. The main components of the proposed model are described as follows.

\textit{1) Convolutional Block:} To extract spatial features, a two-dimensional convolution operation is applied to the traffic flow data $X[t^{s}]$ at time $t^s$. The convolution operation involves a two-dimensional convolution kernel filter, which slides over the flow data to acquire the local perceptual domain. The convolution operation can be expressed as

\begin{equation}\label{Enc_eq}
        Y[t^{s}]=\sigma(W_{s}*X[t^{s}]+b_{s}),
\end{equation}

where $W_{s}$ represents the filter weights, $b_{s}$ is the bias term, $X^{s}[t]$ denotes the input traffic flow at time $t^s$, $\ast$ denotes the convolution operation, $\sigma$ represents the activation function, and $Y[t^{s}]$ is the output of the first convolutional layer. This process helps in extracting spatial features from the neighboring observation locations. $G[t^{s}]$ represents the output of the second convolutional layer.
\begin{equation}\label{Enc_eq}
        G[t^{s}]=\sigma(W_{s_{2}}*Y[t^{s}]+b_{s_{2}})
\end{equation}

After processing the current spatiotemporal information through the two convolutional layers, the output is then connected to the squeeze and excitation module.

\begin{figure}[t]
        \centering
        \includegraphics[width=3.4in]{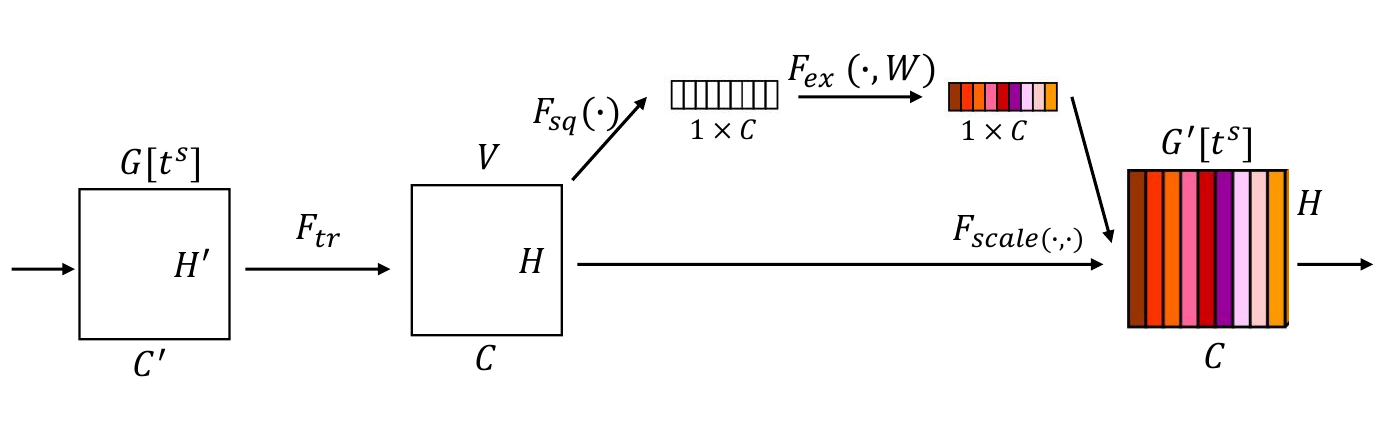}
        \caption{The weighting mechanism within the Squeeze-and-Excitation block.}
        \label{fig:SE}
\end{figure}

\begin{figure}[t]
        \centering
        \includegraphics[width=3.4in]{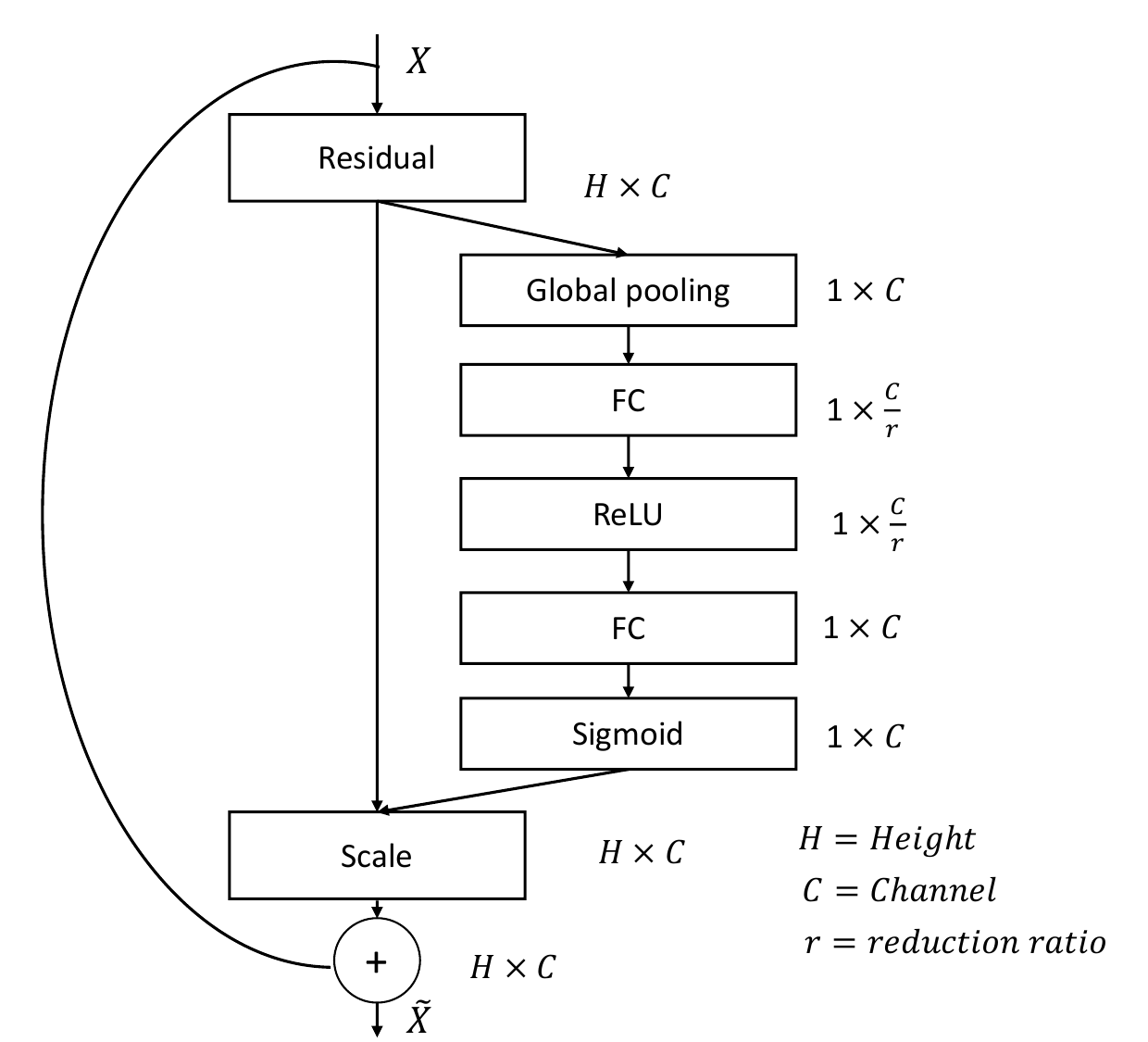}
        \caption{Squeeze-and-Excitation module architecture.}
        \label{fig:SE-res}
\end{figure}

\textit{2) Squeeze-and-Excitation (SE):}
In the SE, convolution transformation is represented by $F_{tr}$, which maps the input $G[t^{s}]$ to feature mappings ${V}$ where ${V} \in \mathbb{R}^{H \times C}$ (see Fig.~\ref{fig:SE}). The feature mappings ${V}$ undergo a squeeze operation, which aggregates the feature maps across their spatial dimensions $(H)$ to generate a channel descriptor. This descriptor captures the global distribution of channel-wise feature responses, allowing all network layers to access information from the entire receptive field. Subsequently, the excitation operation, implemented through a self-gating mechanism, takes the channel descriptor as input and produces modulation weights specific to each channel. These weights are then applied to the feature mappings ${V}$, generating the output of the SE block. This output can be directly fed into subsequent layers of the network. In our model, one dimensional SE is applied to the input $G[t^{s}]$ to generate the output is $G'[t^{s}]$, which is input to the LSTM module. The complete architecture for the SE module is given in Fig. \ref{fig:SE-res}.

\textit{3) LSTM:} Long-term dependencies within sequential data can be efficiently captured using the LSTM architecture, making it particularly suitable for handling extended sequential patterns. In our model, we employ multiple LSTM layers to capture higher-level traffic flow features. The first LSTM  processes the sequence output from the SE module $G^{'}[t^s] = [G'[t^{s}-n],\dots, G'[t^{s}-1], G[t^{s}]]$ and calculates the hidden state for each time step $H_{1}[t^{s}] = [H_{1}[t^{s}-n],..., H_{1}[t^{s}-1], H_{1}[t^{s}]]$.
Then the hidden state sequence $H_{1}[t^{s}]$ is input into the second LSTM layer to calculate the hidden state $H_{2}[t^{s}]$ as the output, which indicates the spatial-temporal feature for time step $t^s$.
LSTM layers are stacked so that each subsequent layer receives the hidden state of the previous layer. As a result, the model can capture increasingly complex patterns and dependencies within the sequential data. The diagram in Fig.~\ref{fig:attention} visually represents the used LSTM layers and their sequential connections.

\textit{4) Attention Mechanism: } The standard LSTM cannot determine the importance of different parts within a traffic flow sequence. To address this limitation, an attention mechanism is introduced. This attention mechanism enables the model to automatically identify varying levels of importance for different segments of the traffic flow sequence at different time steps. The incorporation of the attention mechanism with the Conv-LSTM module is depicted in Fig. \ref{fig:attention}, providing a visual representation of its functionality. The output of Conv-SE-LSTM at each time step $t^s$ is computed as a weighted summation of the output of the LSTM network $H_2[t^{s}]$ follows:

\begin{figure}[t]
        \centering
        \includegraphics[width=4.0in]{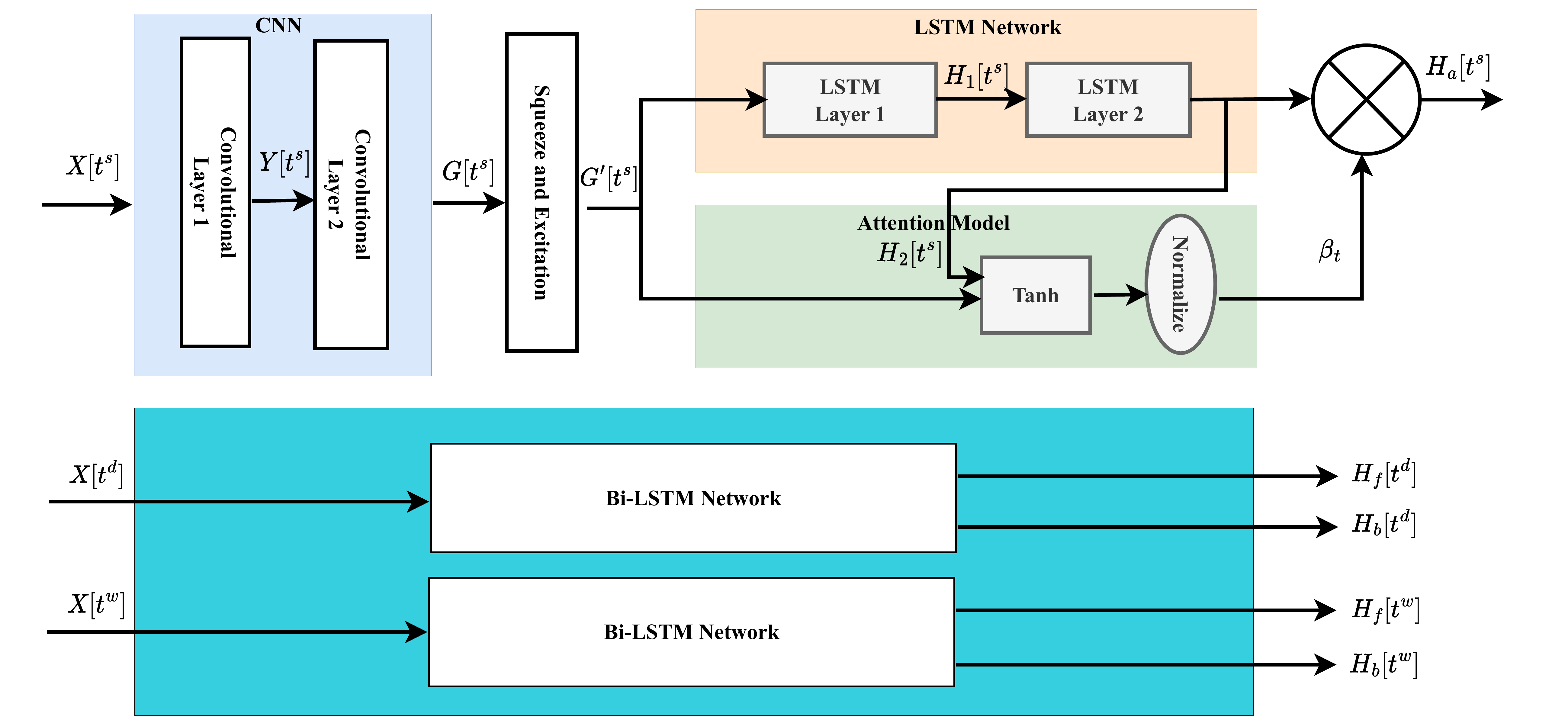}
        \caption{The Conv-SE-LSTM module with an attention mechanism.}
        \label{fig:attention}
\end{figure}

\begin{equation}\label{Enc_eq}
\ H_{a}[t^s]=\sum_{k=1}^{n+1} \beta_{k}H_2[t^{s}-(k-1)]\
 \end{equation}
where $n + 1$ is the length of flow sequence and $\beta_{k}$ is the temporal attention value at time step $t-(k-1)$. The attention value $\beta_{k}$ can be computed as
\begin{equation}\label{Enc_eq}
\ \beta_{k}=\frac{exp(s_{k})}{\sum_{k=1}^{n+1}exp(s_{k})}\
 \end{equation}

The scores $s=(s_{1},s_{2},...,s_{n+1})^T$ indicate the importance of each part in the traffic flow sequence, which can be obtained as
 \begin{equation}\label{Enc_eq}
s_{t}=V_{s}^{T}tanh(W_{hs}G[t^{s}]+W_{ls}H_2[t^{s}])
\end{equation}
where $V_{s}^{T}, W_{hs}$ and $W_{ls}$ are the learnable parameters and $H_2[t^{s}]$ is the hidden output from the Conv-LSTM network.

\begin{figure}[t]
\centering
\includegraphics[width=3.4in]{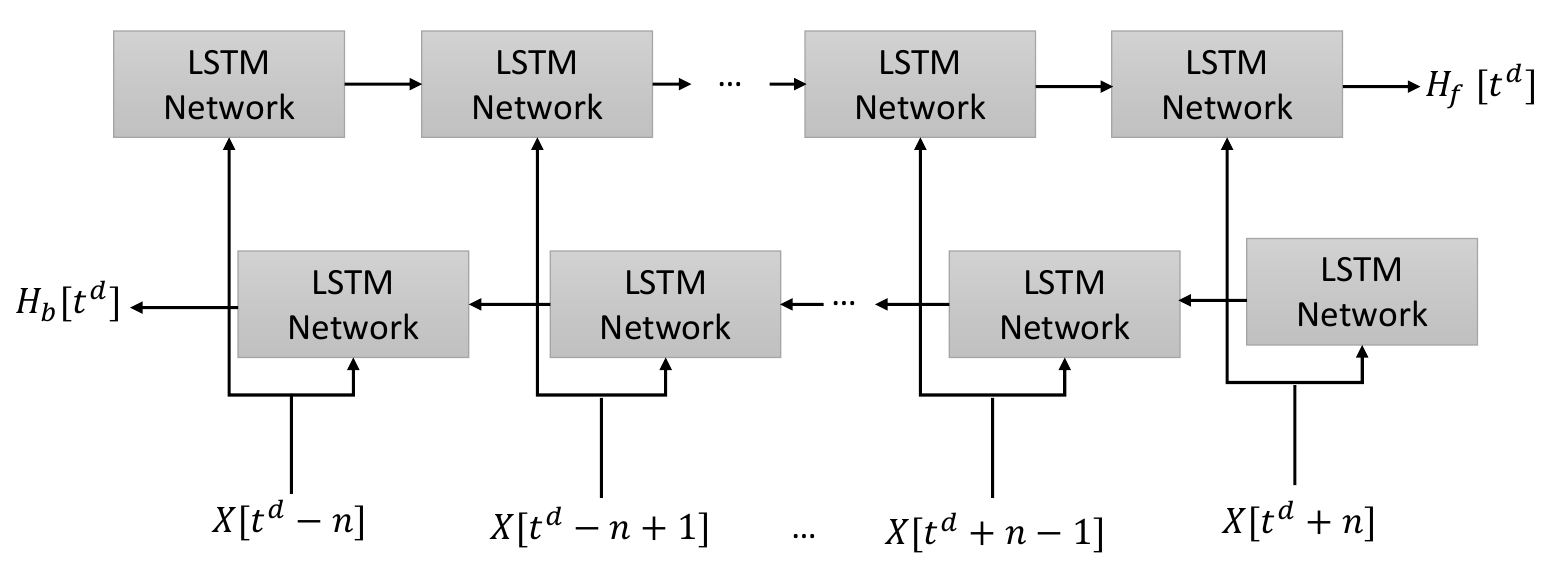}
\caption{The structure of Bi-LSTM networks.}
\label{fig:BI-LSTM}
\end{figure}

\textit{5) Bidirectional LSTM (Bi-LSTM):} A module based on bi-directional LSTM networks is employed to extract periodic features and capture such a temporal dependency from the daily $\textbf{X}[t^d]$ and  $\textbf{X}[t^w]$  weekly densities. The hidden states of forward and backward passes are combined as the output. This way, more features from both directions can be captured, improving the prediction performance. Fig. \ref{fig:BI-LSTM} illustrates the overall structure of the bi-directional LSTM module used in the model.

\begin{figure}[t]
\centering
\includegraphics[width=3.4in]{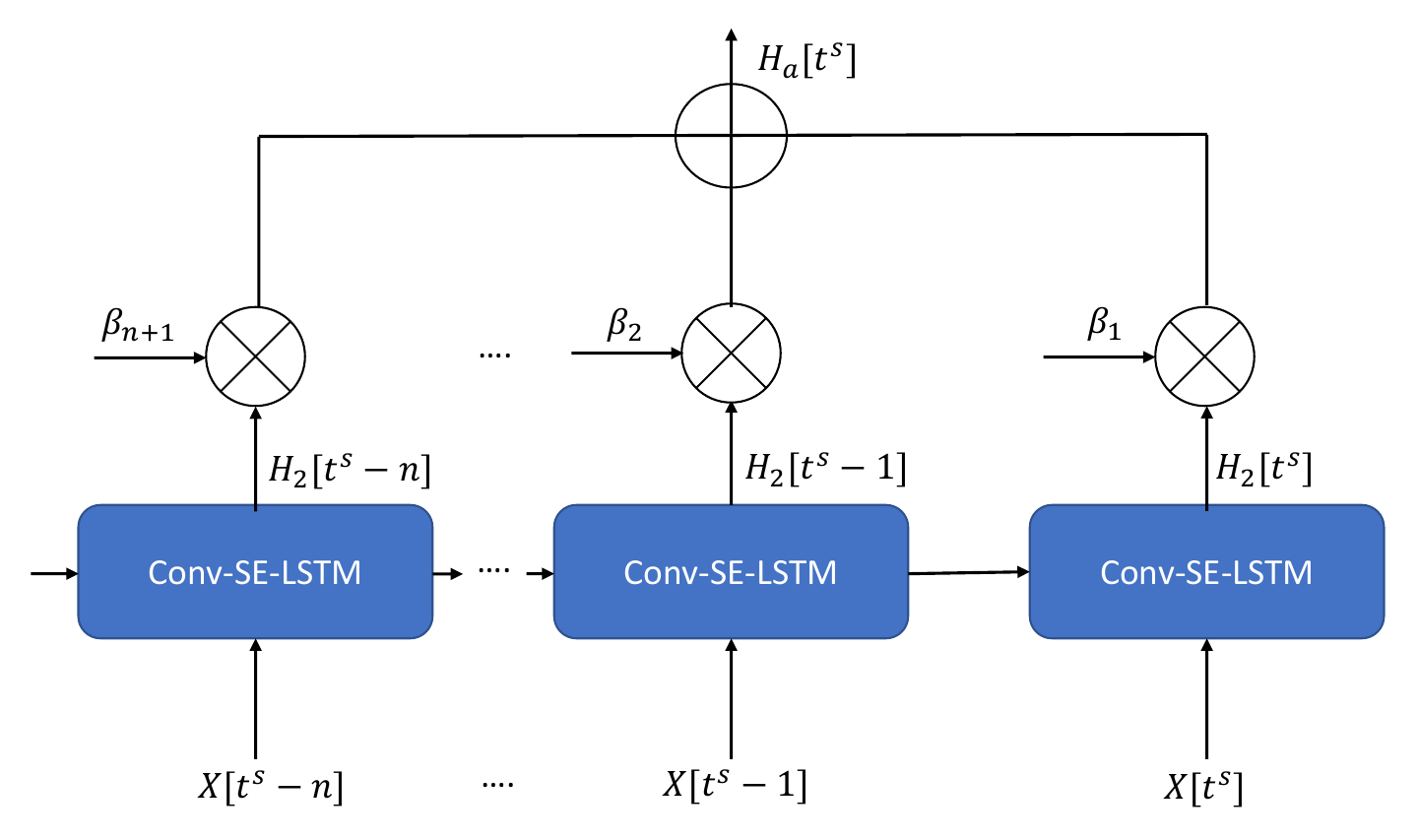}
\caption{The attention mechanism with Conv-LSTM networks.}
\label{fig:at-mecha}
\end{figure}

% As illustrated in Fig. 6, the output of Conv-LSTM at each time step $t$ is computed as a weighted summation of the output of the LSTM network $H_{t}^{s}$ follows:
%        \begin{equation}\label{Enc_eq}
% \ H_{t}^{a}=\sum_{k=1}^{n+1} \beta_{k}H_{t-(k-1)}^{s}\
%         \end{equation}
% where $n + 1$ is the length of flow sequence and $\beta_{k}$ is the temporal attention value at time step $t-(k-1)$. The attention value $\beta_{k}$ can be computed as
%        \begin{equation}\label{Enc_eq}
% \ \beta_{k}=\frac{exp(s_{k})}{\sum_{k=1}^{n+1}exp(s_{k})}\
%         \end{equation}

% The scores $s=(s_{1},s_{2},...,s_{n+1})^T$ indicate the importance of each part in the traffic flow sequence, which can be obtained as
%         \begin{equation}\label{Enc_eq}
% s_{t}=V_{s}^{T}tanh(W_{hs}G_{t}^{s}+W_{ls}H_{t}^{s})
%    \end{equation}
% where $V_{s}^{T}, W_{hs}$ and $W_{ls}$ are the learnable parameters and $H_{t}^{s}$ is the hidden output from the Conv-LSTM network.

% A module based on bi-directional LSTM networks is created to extract periodic features and capture such a temporal dependency from the historical traffic flow. The hidden states of forward and backward passes are combined as the output. This way, more features from both directions can be captured, improving the prediction performance. Fig. 7 illustrates the overall structure of the proposed bi-directional LSTM module used in the model.

As shown in Fig.~\ref{fig:attention}, $H_{t}^{a}$ can be obtained after the processing by the attention Conv-LSTM and Bi-LSTM modules, the spatial-temporal features, the daily periodicity features $H_{t}^{d,f}$, $H_{t}^{d,b}$ the weekly periodicity features $H_{t}^{w,f}$ and $H_{t}^{w,b}$. Then, all these features are concatenated into a feature vector and then input by two regression layers to perform forecasting.

As shown in Fig.~\ref{fig:attention}, the spatial-temporal features $H_{a}[t]$, the daily periodicity features $H_{f}[t^d]$, $H_{b}[t^d]$ and the weekly periodicity features $H_{f}[t^w]$ and $H_{b}[t^w]$ can be obtained after the processing by the attention Conv-SE-LSTM and Bi-LSTM modules. Then, these features are concatenated into a feature vector fed into two regression layers to carry out forecasting.

\textbf{Architecture Remarks.} In our model, we utilize Squeeze-and-Excitation (SE) layers to enhance the performance of convolutional neural networks (CNNs) by adaptively recalibrating the channel-wise feature responses. The SE layer employs global pooling to reduce the spatial dimensions of the input data, generating a channel descriptor for each channel. This descriptor is then processed through a fully connected layer to generate channel weights. These weights are utilized to scale the original feature maps, enabling the network to selectively emphasize different regions of the input data based on the specific task at hand.

The attention mechanism is also employed to selectively focus on specific segments of the input data rather than processing the entire input indiscriminately. Attention is commonly used in sequence-to-sequence models like Recurrent Neural Networks (RNNs) and Transformer-based models, particularly when dealing with variable-length input sequences. The model can assign weights to different parts of the sequence by employing attention mechanisms based on their relative importance for the given task. This allows the model to effectively allocate its attention and resources to the most relevant portions of the input sequence.

\section{Privacy and Security Analysis}

% \hlg{Building it (Dr. Mahmoud).}

\textcolor{black}{    \hspace{5mm}\textit{\textbf{Proposition 1}.
    Insider and outsider adversaries cannot gain access to the drivers' locations.
    } }

    \vspace{1mm}
    \textcolor{black}{\textit{\textbf{Proof.}}
    In our scheme, we employ an inner product encryption\mbox{\cite{boneh2011functional}} cryptosystem to protect the drivers' reported locations. This encryption technique ensures that without knowledge of the secret keys, decrypting the ciphertexts is impossible. Since each driver is assigned a unique key, and only the TMC (Traffic Management Center) possesses the decryption key, it is infeasible to decrypt the ciphertexts using other drivers' secret keys as well. For the TMC, it receives $K$ ciphertexts from each driver, each has an encrypted value of either $0$ or $1$. It can be shown that there is a negligible computational probability of distinguishing between these encrypted values due to $K$ anonymity.}

\vspace{2mm}
\textcolor{black}{	\hspace{5mm}\textit{\textbf{Proposition 2}.
	The encrypted cells of the same drivers are not linkable under the known ciphertext model.
	} }

    \vspace{1mm}
    \textcolor{black}{\textit{\textbf{Proof.}}
    During the encryption process, the IPFE\mbox{\cite{boneh2011functional}} employs a random nonce to ensure that ciphertexts originating from the same cells and the same driver appear distinct and remain unlinkable. To be specific, for each driver's $i$ encrypted cell $j$, the driver generates a unique random number by selecting a random element $r_i^j \leftarrow_R \mathbb{Z}_p$ for $j \in {1, \ldots, K}$ and uses this value in the encryption process. Furthermore, the driver employs K-anonymity\mbox{\cite{sweeney2002k}} to report $K-1$ other random encrypted cells to obscure their true location. These features are crucial in preventing the linking of ciphertexts corresponding to the same driver who visits different cells. Prolonged tracking of a driver's location cannot lead to identifying that person based on the visited locations. We conclude that our scheme is secure in the \textit{known ciphertext model}, where attackers cannot obtain the secret keys or the plaintext information using the encrypted cells.}

\vspace{2mm}
 \textcolor{black}{   \hspace{5mm}\textit{\textbf{Proposition 3}.
   Each driver is unable to decrypt the ciphertexts of other drivers since a shared key is not employed; rather, each driver possesses a unique secret key  } }

    \vspace{1mm}
  \textcolor{black}{  \textit{\textbf{Proof.}}
    If a driver could decrypt the ciphertexts belonging to other drivers, it would jeopardize the privacy of locations visited by those drivers. In our scheme, the encrypted cells generated by one driver remain inaccessible to others, primarily because each driver utilizes a distinct key denoted as $\mathbf{pk}_i$\mbox{\cite{boneh2011functional}}. Despite using different keys by drivers when reporting their encrypted cells, the TMC can still assess the density of drivers within each cell by calculating the inner product of the reported cell encryptions for cell $j$ and the vector $\textbf{y}$ consisting of ones\mbox{\cite{boneh2011functional}}.
    }

% 		\vspace{2mm}
% \textcolor{black}{	\hspace{5mm}\textit{\textbf{Preposition 4}.
% 	The TMC should not be able to match a large number of indices and trapdoors to avoid leaking side information.}}

%     \vspace{2mm}
%   \textcolor{black}{  \textit{\textbf{Proof.}}
    %The cloud server should be able to match indices and trapdoors to find the locations visited by persons of interest without being able to identify the persons. However, suppose the cloud server has a large amount of data collected over a long time. In that case, it may use the data to infer statistical and side information, such as collecting many locations visited by an anonymous person of interest.
   % To prevent the cloud server from collecting side information, the keys of the involved parties should change frequently, e.g., every month, to make sure that the ciphertexts sent after updating the keys cannot be matched to the old ciphertexts because they are encrypted with different keys.
    % }

\section{Performance Analysis} \label{sec:Performance Evaluations}
\begin{figure}[t]
    \captionsetup{labelfont=it,justification=centering}
    \centering
       \includegraphics[width=3.5in, height=2in]{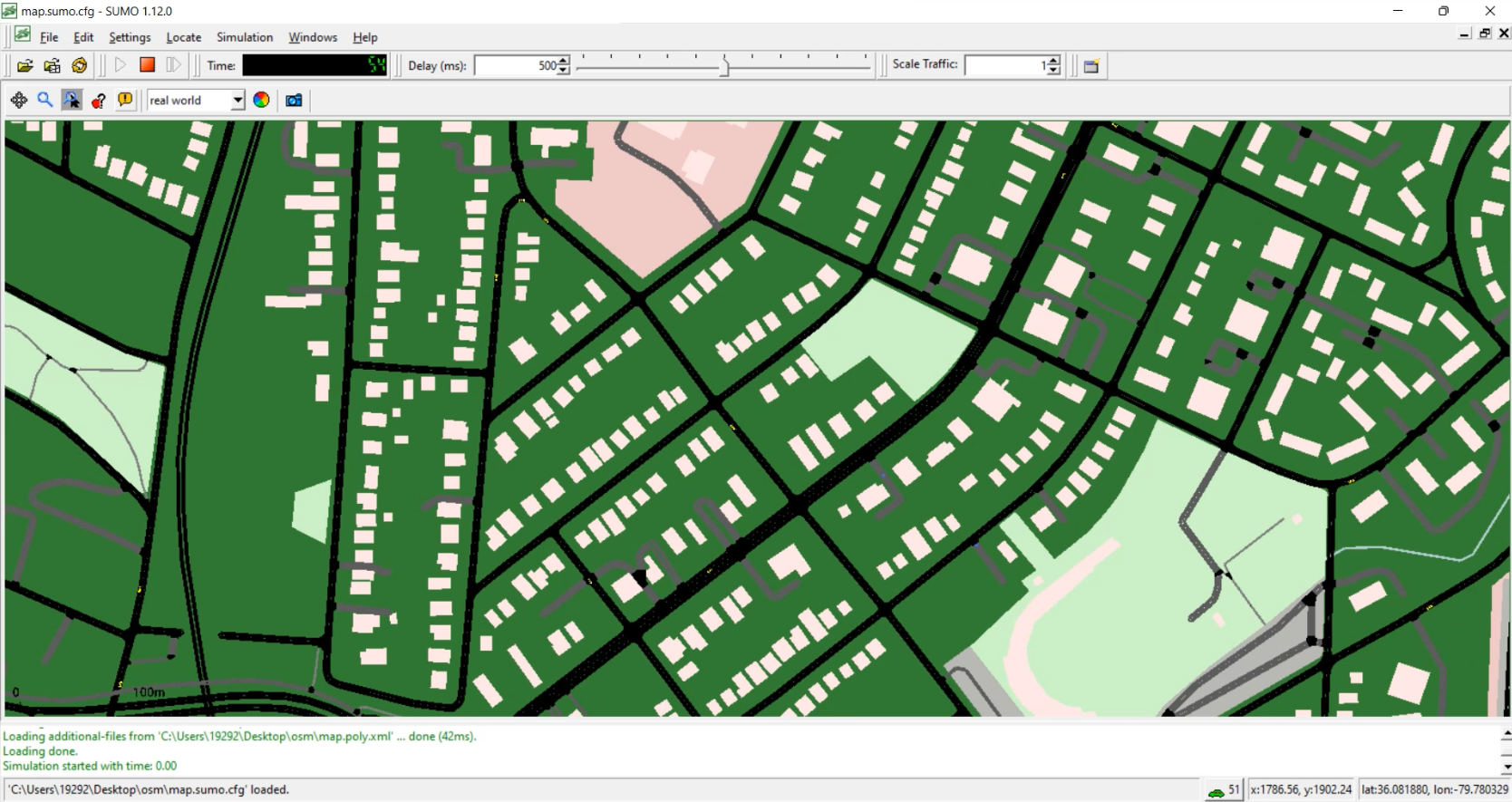}
       \caption{Synthetic dataset generation using SUMO}
       \label{fig:sumo}
\end{figure}

The proposed schemes were implemented in Python on a Lambda GPU workstation equipped with the following specifications: 2xQuadro RTX 8000 GPUs, 2-Way NVLink, Intel i9-9820X CPU (10 Cores), 128 GB of RAM, and a 2 TB NVMe SSD. This workstation came pre-installed with the latest versions of essential libraries such as CUDA, Jupyter, Pytorch, Tensorflow, and Keras. For our implementation, we utilized two datasets:

\begin{itemize}
    \item \textbf{SUMO Dataset:} To assess the encryption component of our project, we generated a set of random trips based on real maps. We started by obtaining a genuine map of Greensboro, North Carolina, USA, from the OpenStreetMap project\mbox{\cite{OpenStreetMap}}. The traffic management area covered an $8\, \text{km} \times 8\, \text{km}$ region, divided into 40 cells, each measuring $1\, \text{km} \times 1\, \text{km}$. To create real and random routes, we employed the "Simulation of Urban MObility" (SUMO) software\mbox{\cite{ITSBerlin2015}}. All results presented are the averages from 30 different runs (See Fig.~\ref{fig:sumo}).\\

    \item \textbf{PeMS Dataset:} This dataset was sourced from the Performance Measurement System (PeMS), supported by California Department of Transportation (Caltrans)\mbox{\cite{PeMS}}. We used the PeMS14 dataset, covering traffic data from 2001 to 2023 across California's major metropolitan areas. The data, collected from nearly 40,000 sensors, is mostly recorded at 5-minute intervals, with some available at 30-second intervals for more detailed historical and real-time traffic analysis. For our study we focused on two specific scenarios: freeway and urban traffic, training and evaluating our proposed model with data from 183 sensors in District 10, specifically on Freeway SR99-S, as well as 12 sensors from District 4 on Street I980 in Oakland. This enabled robust analysis across both freeway and urban traffic conditions.
\end{itemize}
\vspace{6pt}
We then assess the proposed privacy-preserving traffic management forecasting system from three perspectives: Computation Overhead, Communication Overhead, and Traffic Flow Forecasting.

\begin{figure*}[t]
        \begin{subfigure}{0.48\textwidth}
            \centering
            \includegraphics[width=3.5in, height=2.5in]{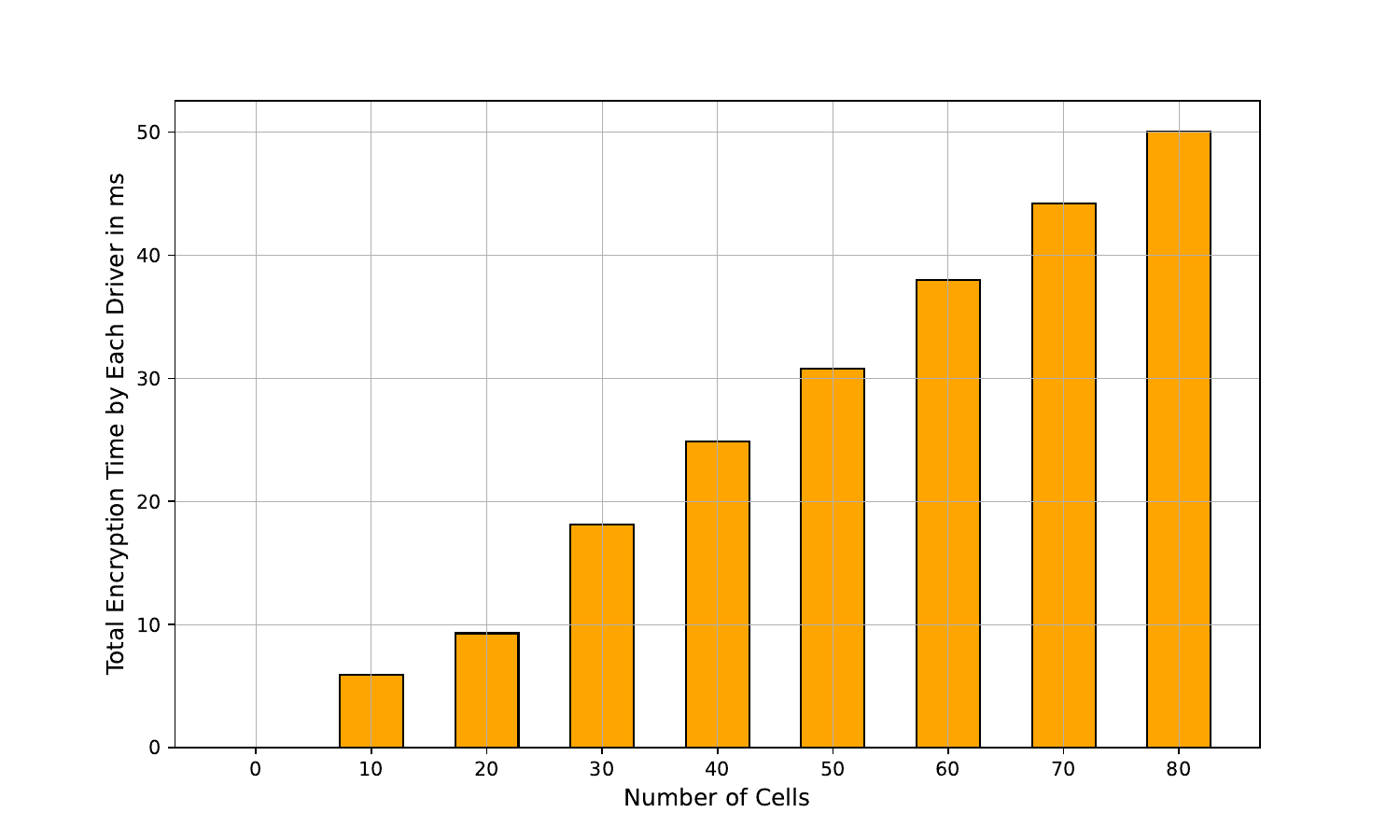}
            \caption{Encryption times vs Number of Cells.}
            \label{fig:et}
        \end{subfigure}
        \hfill
        \begin{subfigure}{0.48\textwidth}
            \centering
            \includegraphics[width=3.5in, height=2.5in]{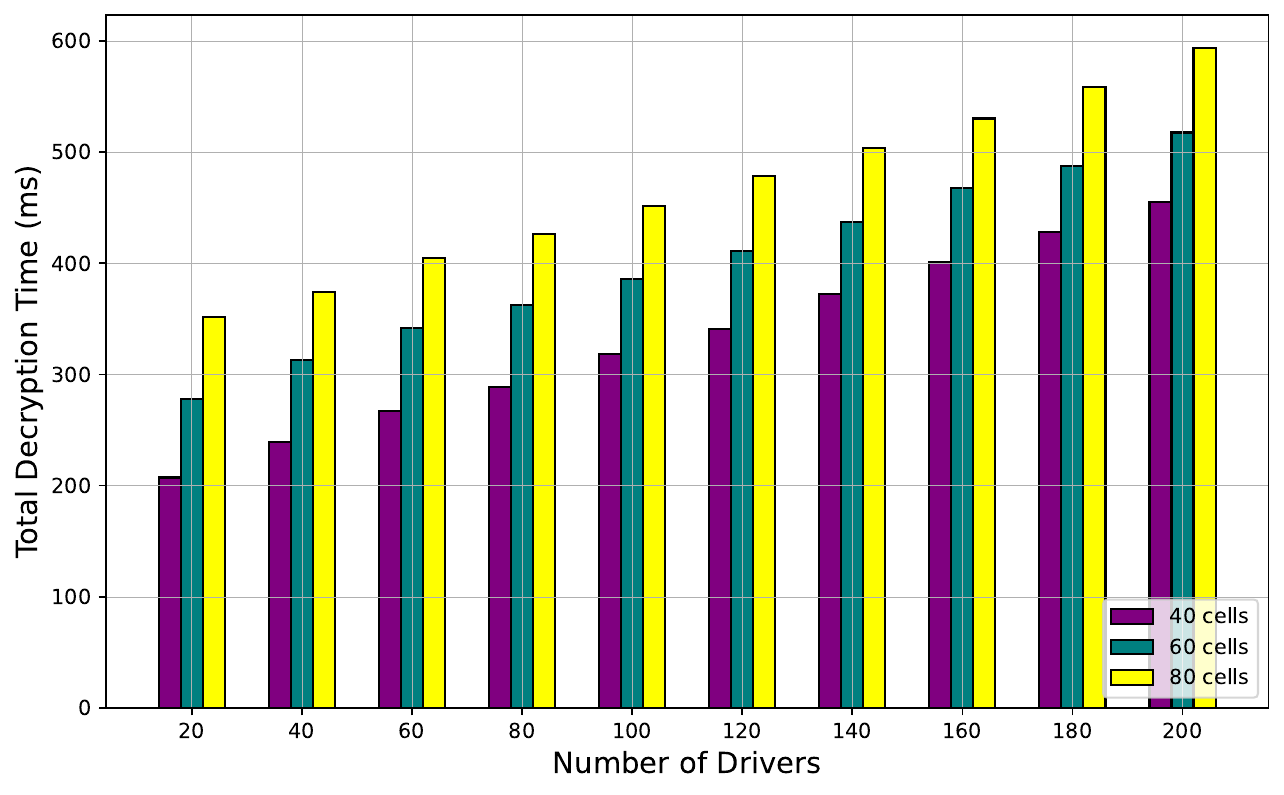}
            \caption{Decryption Time vs Number of Drivers.}
            \label{fig:dt}
        \end{subfigure}
        \caption{Computation Overhead Analysis.s.}
        \label{performance_results}
\end{figure*}

\subsection{Computation Overhead}
The total computation overhead is quantified through two key metrics: the number of cryptographic keys distributed to the drivers from the TMC and the size of the encrypted messages transmitted to the TMC $(D_k + E_m)$. For the first metric ($D_k$), each driver is assigned a key denoted as $\mathbf{pk}_i:= \left(\mathcal{G}, \left[\mathbf{a}_i\right], \left[\mathbf{W}_i \mathbf{a}_i\right], u_i\right)$. Using the asymmetric pairing curve BN256 with a size of 256 bits, where each group element occupies 32 Bytes, in addition to the public parameters, each driver requires two group points (64 bytes each) and one field element (2 bytes in size) to encrypt its location report. The total key size sent from the TMC  ($D_k$), including the public parameters, is 66 bytes. For the second metric ($E_m$), when a driver encrypts a single cell and employs the asymmetric pairing curve BN256 with a size of 256 bits, where each group element is 32 Bytes, the resulting encrypted cell size is composed of two group elements, totaling 64 Bytes. As drivers employ K-anonymity\mbox{\cite{sweeney2002k}} when reporting their locations, the overall message size is directly proportional to the value of `k', where the size of message sent is (64 bytes $\times$ k) with `k' being the number cells to be encrypted by the driver. Thus, for an 80-cell geographical area, drivers require only 5.12 kilobytes of computational resources to encrypt their route data prior to transmission to the TMC. This minimal overhead highlights the efficiency of our scheme, ensuring that even resource-constrained vehicle modules can perform encryption seamlessly, enabling secure and rapid data transmission through existing communication protocols.

\subsection{Communication Overhead}
On the driver's side, overhead is associated with the time it takes them to calculate, encrypt, and transmit their location data. Similarly, the time the Traffic Management Center (TMC) spent to receive, decrypt, and consolidate location reports also contributes to overhead. The randomly generated traffic location data for the Greensboro city area in North Carolina is encrypted and decrypted using the In-Place Functional Encryption (IPFE) technique\mbox{\cite{boneh2011functional}} implemented in the Go programming language. We conducted a simulation in which 100 drivers (represented as $D$ = 200) entered various cells. Figure \mbox{\ref{fig:et}} depicts the relationship between the number of cells that a driver needs to encrypt and the total encryption time in milliseconds. The figure illustrates a direct proportional relationship between the encryption time and the number of cells within the millisecond range, which represents a relatively small privacy cost of approximately 50ms for encrypting an area with 80 cells. Figure \mbox{\ref{fig:dt}} illustrates the relationship between millisecond decryption time and the number of active drivers entering specific cells. Notably, for a scenario involving 40 drivers, 60 drivers and 80 cells, the decryption process typically requires less than 600 milliseconds for an 80-cell area accommodating 200 driver presence. This number underscores the efficiency of the decryption process for such configurations.

    {\renewcommand{\arraystretch}{1.35}
    \begin{table}[t]
    \centering
    \caption{Hyper-parameter tuning}
    \label{tab:Hyper-para}
    \begin{tabular}{| >{\centering\arraybackslash}m{2.2cm}|>{\centering\arraybackslash}m{2.2cm}| >{\centering\arraybackslash}m{2.2cm}|}
    \hline \rowcolor[gray]{0.8}
    \textbf{Hyper-parameter} & \cellcolor[gray]{0.8}  \textbf{Value} & \cellcolor[gray]{0.8}  \textbf{Selected Best Value} \\ \hline
     Units      &  32, $\dots$, 512 & 488 \\ \hline
     Activation   & \texttt{relu, tanh, sigmoid}   & \texttt{relu} \\ \hline
     Dropout      &  True, False  &  True  \\ \hline
     Learning rate      &  $1 \times 10^{-4}$ to $1 \times 10^{-2}$  &  0.0003  \\ \hline
     \end{tabular}
    \end{table}}

    {\renewcommand{\arraystretch}{1.35}
    \begin{table}[t]
    \centering
    \caption{Different Optimizer comparison}
    \label{tab:Different Opt}
    \begin{tabular}{| >{\centering\arraybackslash}m{2.2cm}|>{\centering\arraybackslash}m{1.2cm}| >{\centering\arraybackslash}m{1.2cm}| >{\centering\arraybackslash}m{1.2cm}|}
    \hline \rowcolor[gray]{0.8}
    \textbf{Optimizer} & \cellcolor[gray]{0.8}  \textbf{MAE} & \cellcolor[gray]{0.8}  \textbf{MAPE} & \cellcolor[gray]{0.8}  \textbf{RMSE} \\ \hline
    \texttt{SGD}      &  39.45    &  62.73 & 45.66\\ \hline
    \texttt{ADADELTA}   & 18.75   & 19.13 & 24.09 \\ \hline
    \texttt{RMSProp}      &  12.06   &  14.25 & 15.67  \\ \hline
    \texttt{ADAGRAD}      &  9.80   &  10.18 & 13.93  \\ \hline
    \texttt{ADAM}      &  7.94   &  8.50 & 11.03 \\ \hline
    % $h_5$      &  350   &  ReLU \\ \hline
    % $h_6$      &  110   &  ReLU \\ \hline
    % $h_7$      &  1536  &  ReLU \\ \hline
    % $h_8$      &  500   &  ReLU \\ \hline
    % $h_9$      &  1536  &  ReLU \\ \hline
    % $h_{10}$   & 500    &  ReLU \\ \hline
    % $h_{11}$   & 1536   &  ReLU \\ \hline
    % $h_{12}$   & 500    &  ReLU \\ \hline
    % $h_{13}$   & 700    &  ReLU \\ \hline
    % Output     & 2      &  Softmax\\ \hline
    \end{tabular}
    \end{table}}

     \begin{figure*}[!ht]
        \begin{subfigure}{0.48\textwidth}
            \centering
            \includegraphics[width=\linewidth, height=2.5in]{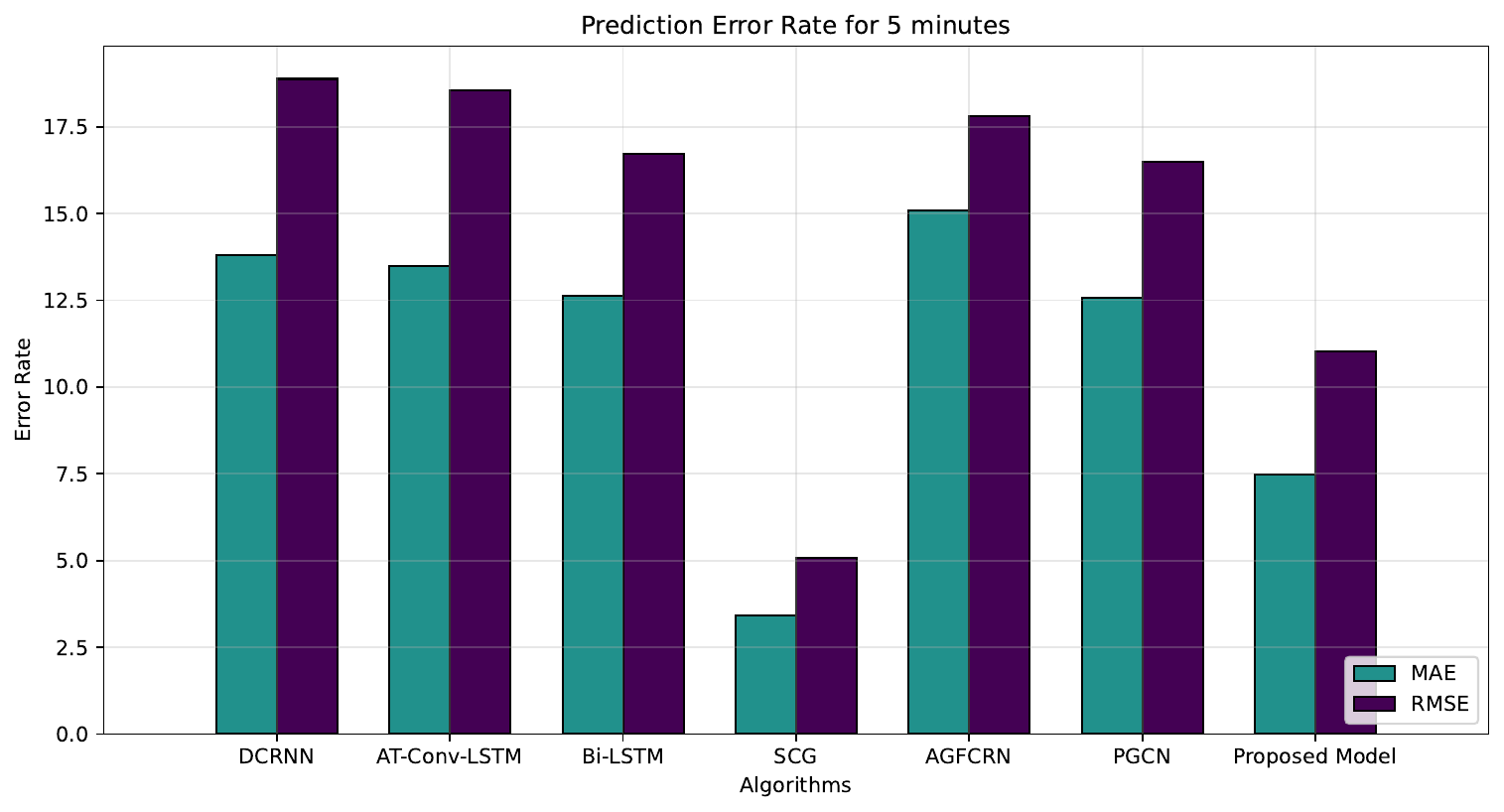}
            \caption{Mean Absolute Error and Root Mean Square Error}
            \label{mse_rmsea}
        \end{subfigure}
        \hfill
        \begin{subfigure}{0.48\textwidth}
            \centering
            \includegraphics[width=\linewidth, height=2.5in]{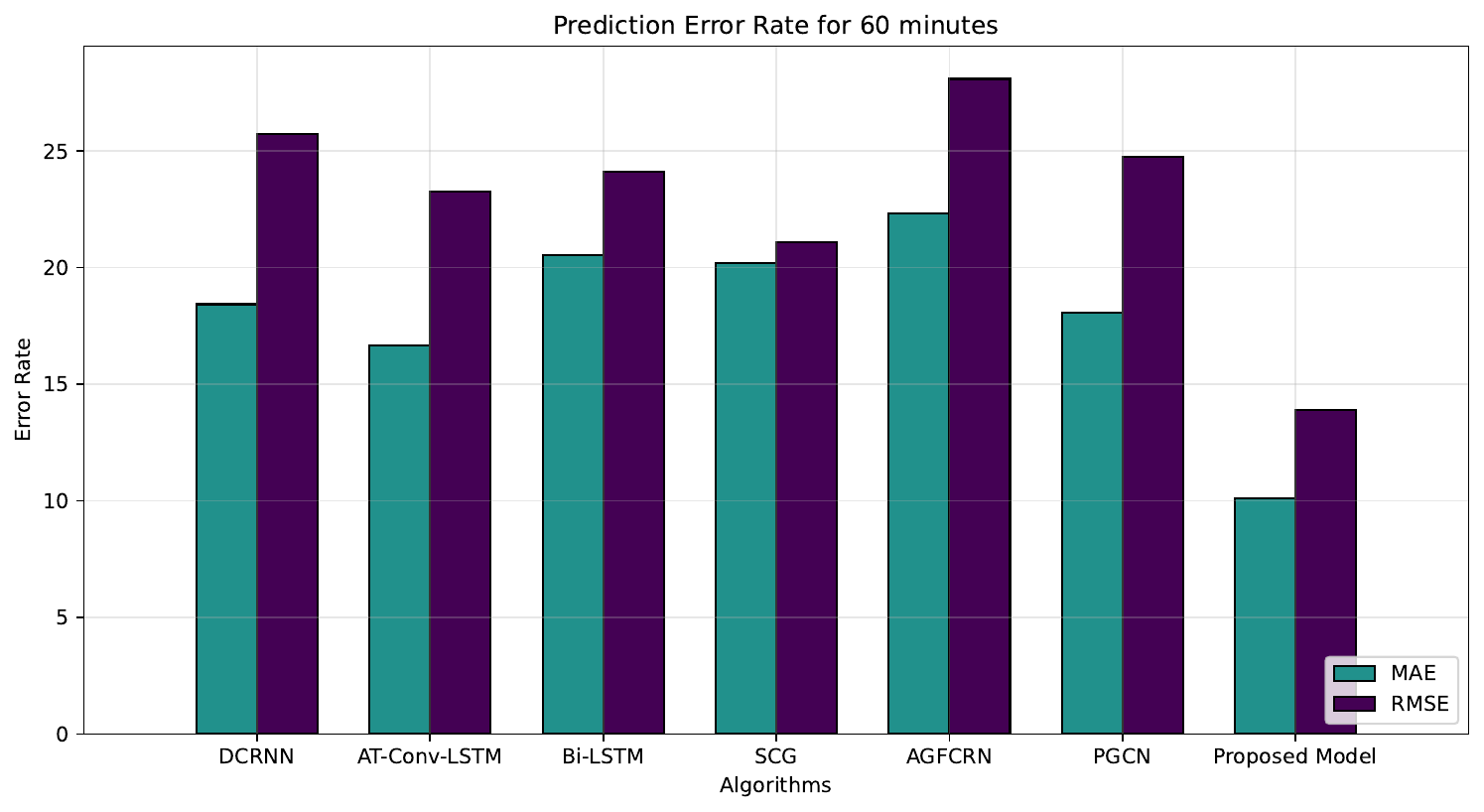}
            \caption{Mean Absolute Error and Root Mean Square Error}
            \label{mse_rmseb}
        \end{subfigure}
        \caption{Error Rate Assessment for Short-Term Traffic Flow Forecasting.}
        \label{performance_results}
    \end{figure*}

    \begin{figure}[!h]
    \centering
    \includegraphics[width=3.8in]{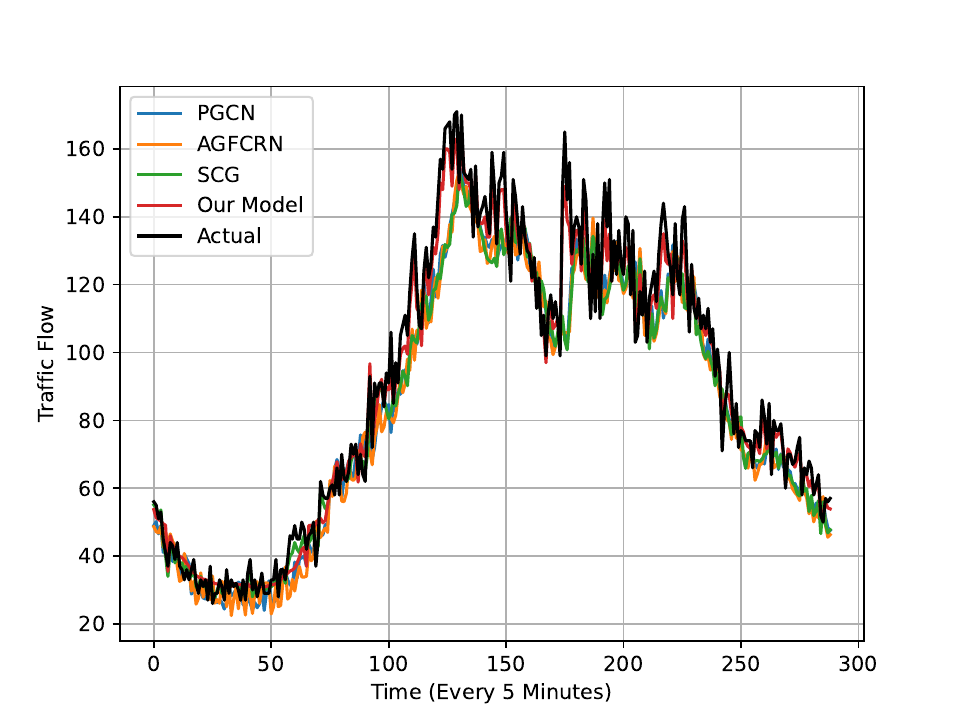}
    \caption{Comparison of Model Predictions and Actual Traffic Flow over a 300-Minute Interval.}
    \label{traffic_flow}
    \end{figure}
\subsection{Traffic Flow Forecast}
This subsection evaluates the traffic forecasting model (Conv-LSTM) both with and without the squeezing and excitation algorithms, attention mechanism, and Bi-LSTM. In order to measure our suggested scheme against comparable traffic forecasting methods found in the literature, we selected three commonly used performance indices. These measures, \texttt{Mean Absolute Error (MAE)}, \texttt{Mean Absolute Percentage Error (MAPE)}, and \texttt{Root Mean Square Error (RMSE)} assess the accuracy of predictive models in regression analysis.\\
\begin{itemize}
        \item \textit{Mean Absolute Error:} The MAE is calculated using the formula
        \begin{equation}\label{Enc_eq}
            \ MAE=\frac{1} {n} \sum_{t=1}^{n} |F_{p}-F_{t}|\
        \end{equation}
        \vspace{6pt}
        \ \item \textit{Mean Absolute Percentage Error:} The MAPE is calculated as follows: \begin{equation}\label{Enc_eq}
            \ MAPE (\%) =\frac{1} {n} \sum_{t=1}^{n} |\frac{F_{p}-F_{t}} {F_{t}}| \times 100\
                    \end{equation}
        \vspace{3pt}
        \item \textit{Root Mean Square Error:} The RMSE is determined by the formula: \begin{equation}\label{Enc_eq}
            \ RMSE=\sqrt{\frac{1} {n} \sum_{t=1}^{n} (F_{p}-F_{t})^{2}}\
                    \end{equation}
    \end{itemize}
where $F_{p}$ represents the predicted traffic flow and $F_{t}$ represents the true traffic flow.\\

\begin{enumerate}
    \item \textbf{Experimental Data and Evaluation:}
    Using the PeMS dataset, the hyperparameters of the forecasting Conv-LSTM model are fine-tuned utilizing the Tensorflow Keras Tuner. The tuning process involved exploring a range of hyperparameter values, including unit limits for feed-forward layers ranging from 32 to a maximum of 512, likewise exploring various activation functions ranging \texttt{ReLU}, \texttt{Sigmoid}, and \texttt{Tanh}. Lastly, we investigated different learning rates within $1 \times 10^{-4}$ to $1 \times 10^{-2}$. Table \ref{tab:Hyper-para} contains the tuning process outcomes and the optimal hyperparameter values. Additionally, we performed a comparative analysis on adopting five different optimizers for our model. The optimizers used were Stochastic Gradient Descent (SGD), ADADELTA, Root Mean Square Propagation (RMSProp), Adaptive Gradient (ADAGRAD), and Adaptive Moment Estimation (ADAM). The analysis results are presented in Table \ref{tab:Different Opt}, where the ADAM consistently outperforms other optimizers regarding error reduction. Consequently, we selected ADAM as the optimizer for our final model.\\

{\renewcommand{\arraystretch}{1.5}
\begin{table*}[t]
    \centering
    \caption{Prediction performance with various proposed modules for prediction of urban area traffic flow.} %\hlr{Having difficulties positioning this table properly before the Conclusion}
    \label{tab:Prediction}
\begin{tabular}{|c|c|c|c|c|c|}
\hline
\cellcolor[gray]{0.8}Algorithm &
\cellcolor[gray]{0.8} Measure &
\cellcolor[gray]{0.8}$\mathbf{5} \mathbf{m i n}$ & \cellcolor[gray]{0.8}$\mathbf{1 5} \mathbf{m i n}$ & \cellcolor[gray]{0.8}$\mathbf{3 0} \mathbf{~ m i n}$ & \cellcolor[gray]{0.8}$\mathbf{6 0} \mathbf{~ m i n}$ \\

\hline \multirow{3}{*}{ Conv LSTM (Stage 1)} & MAE & $9.05$ & $10.80$ & $10.28$ & $10.50$ \\
\cline { 2 - 6 } & MAPE (\%) & $9.94 $ & $11.73 $ & $10.25 $ & $11.33 $ \\
\cline { 2 - 6 } & RMSE & $12.32$ & $14.73$ & $14.28$ & $14.21$ \\
\hline \multirow{3}{*}{ Bi-Conv LSTM (Stage 2) } & MAE & $12.57$ & $18.07$ & $12.37$ & $14.31$ \\
\cline { 2 - 6 } & MAPE (\%) & $13.18 $ & $19.08 $ & $12.9 $ & $13.73 $ \\
\cline { 2 - 6 } & RMSE & $18.8$ & $24.16$ & $18.55$ & $21.16$ \\
\hline \multirow{3}{*}{ AT-Bi-Conv LSTM (Stage 3) } & MAE & $8.19$ & $9.45$ & $9.21$ & 10 \\
\cline { 2 - 6 } & MAPE (\%) & $8.86 $ & $9.56 $ & $9.60 $ & $10.71 $ \\
\cline { 2 - 6 } & RMSE & $11.33$ & $13.14$ & $13.01$ & $13.94$ \\
\hline \multirow{3}{*}{ AT-Bi-Conv-SE LSTM (Stage 4) } & MAE & $\mathbf{7 . 9 4}$ & $\mathbf{8 . 6 6}$ & $\mathbf{9 . 8 8}$ & $\mathbf{1 0 . 1 0}$ \\
\cline { 2 - 6 } & MAPE (\%) & $\mathbf{8 . 5 }$ & $\mathbf{9 . 2 2 }$ & $\mathbf{1 0 . 6 6 }$ & $\mathbf{1 0 . 7 5 }$ \\
\cline { 2 - 6 } & RMSE & $\mathbf{1 1 . 0 3}$ & $\mathbf{1 2 . 1 0}$ & $\mathbf{1 3 . 8}$ & $\mathbf{1 3 . 9}$ \\
\hline
\end{tabular}
\end{table*}}

{\renewcommand{\arraystretch}{1.5}
\begin{table*}[t]
    \centering
    \caption{Performance comparison of different Models for Urban traffic flow prediction.}
    \label{tab:Performance}
\begin{tabular}{|c|c|c|c|c|c|c|c|c|}
\hline
\cellcolor[gray]{0.8}Horizon &
\cellcolor[gray]{0.8}Measure &
% \cellcolor[gray]{0.8} LSTM\cite{zhao2017lstm,tian2015predicting} &
\cellcolor[gray]{0.8} DCRNN\cite{li2017diffusion} &
% \cellcolor[gray]{0.8} DNN-BTF\cite{wu2018hybrid} &
\cellcolor[gray]{0.8} AT-Conv-LSTM\cite{zheng2020hybrid} & 
\cellcolor[gray]{0.8} Bi-LSTM\cite{ma2021short} &
\cellcolor[gray]{0.8} SCG\cite{ma2022novel} &
\cellcolor[gray]{0.8} AGFCRN\cite{li2022adaptive} &
\cellcolor[gray]{0.8} PGCN\cite{shin2024pgcn} &
\cellcolor[gray]{0.8} Our Model
\\

\hline \multirow{3}{*}{ 5 min } & MAE  &  13.79  & 13.49 &  12.63  &  \color{black}{\textbf{3.43}}  &  15.10  &  12.56  &  \textbf{7.49} \\
\cline { 2 - 9 } & MAPE (\%)  &  10.7  & 10.1 &  10.49  & 8.60  & 9.67  & 8.74  & \textbf{8.5} \\
\cline { 2 - 9 } & RMSE  & 18.88  & 18.56 &  16.72  &  \color{black}{\textbf{5.09}}  &  17.81  &  16.49 & \textbf{11.03} \\
\hline \multirow{3}{*}{ 15 min } & MAE &  14.79  & 14.34 &  15.09  &  \color{black}{\textbf{6.89}}   &  16.71  &  13.43  &  \textbf{8.66} \\
\cline { 2 - 9 } & MAPE (\%)  &  11.5  &  10.8  &  12.28  & 11.61  &  10.14  &  9.88  &  \textbf{9.22} \\
\cline { 2 - 9 } & RMSE & 20.43  & 20.08 &  18.34  & \color{black}{\textbf{8.43}} &  22.68  &  17.91  & \textbf{12.10} \\
\hline \multirow{3}{*}{ 30 min } & MAE  & 16.05  & 15.48 &  17.41  &  11.15    &  19.53   & 15.62 &  \textbf{9.88} \\
\cline { 2 - 9 } & MAPE (\%)  &  12.4  &  11.4  &  14.5  & 16.89  & 12.82  & 11.61  &  \textbf{10.66} \\
\cline { 2 - 9 } & RMSE  &  21.18  & 21.26  &  19.72  &  14.71  &  25.94  &  19.33  &  \textbf{13.8} \\
\hline \multirow{3}{*}{ 60 min } & MAE  &  18.43  &  16.65  &  20.53  &  20.21  &  22.31  & 18.06 &  \textbf{10.10} \\
\cline { 2 - 9 } & MAPE (\%) &  14.2  & 12.3 &  16.9  &  17.36  &  15.53   &  13.89   &  \textbf{10.75} \\
\cline { 2 - 9 } & RMSE  &  25.74  &  23.26  &  24.12  &  21.09  &  28.09   &  24.74   &  \textbf{13.9} \\
\hline
\end{tabular}
\end{table*}}

    % \hl{I worked on the forecast performance evaluation section, likewise Table V(where I compared our work with current traffic forecasting literature}
    \item \textbf{Forecast Performance Evaluation:}
    Here, we demonstrate the efficacy of our proposed hybrid model for traffic flow prediction at a particular Point of Interest (POI) on Street I980 in Oakland, District 4, by utilizing a number of crucial elements, including an attention mechanism (AT), a squeeze-and-excitation (SE) module, and a Bi-LSTM module. Our hybrid model was built in four stages using the TensorFlow framework\mbox{\cite{girija2016tensorflow}}, Beginning with the Conv-LSTM (Conv LSTM) model (Stage 1). Then, integrating the Conv LSTM model with a Bi-LSTM module to form a Bi-Conv LSTM model (Stage 2). The Bi-Conv LSTM model was enhanced further by adding an attention mechanism to produce an AT-Bi-Conv LSTM model (Stage 3). Lastly, we fuse a squeeze-and-excitation module to the previous AT-Bi-Conv LSTM model, resulting in an AT-Bi-Conv-SE LSTM model (Stage 4). The outcomes, as detailed in Table \mbox{\ref{tab:Prediction}}, highlight our final hybrid model (AT-Bi-Conv-SE LSTM model) as the best performing model, achieving the lowest MAE and RMSE error rates of $7.94 \%$ and $11.03 \%$ respectively for a 5 minutes prediction time, while posing a $10.1 \%$ MAE  value and $13.9 \%$ RMSE value for a prediction horizon of 60 minutes. Also, Table \mbox{\ref{tab:Prediction}} shows stage two (Bi-Conv LSTM model) as the worst performing stage, with the highest MAE and RMSE error rates of $12.57 \%$ and $18.8 \%$ respectively for a 5 minutes prediction horizon, as well as, $14.31 \%$ MAE value and $21.16 \%$ RMSE value for a prediction time of 60 minutes, indicating a decrease in performance, after the addition of the Bi-LSTM module. For instance, a significant increase in MAE and RMSE error rates from $10.5 \%$ and $14.21 \%$ respectively in stage 1 to MAE and RMSE error rates of $14.31 \%$ and $21.16 \%$ respectively in stage 2. Conversely, a substantial performance improvement was witnessed from stage 2 to stage 3, likewise from stage 3 to stage 4 (best performing model) across all prediction horizons. Concretely, this comprehensive approach underscores the effectiveness of our model in accurately forecasting traffic flow and positions it as a leading solution for traffic management and analysis. Fig.\mbox{~\ref{traffic_flow}} shows the prediction performance of the proposed model, emphasizing its superior forecasting accuracy due to its coherence with the referenced actual traffic flow compared to the flow predictions of other baselines forecasting models.

    Furthermore, we comprehensively compared our proposed hybrid (AT-Bi-Conv-SE-LSTM) model and other established contemporary approaches for short-term traffic flow predictions spanning various prediction time horizons (5 minutes, 15 minutes, 30 minutes, and 60 minutes). The comparative approaches encompass Diffusion Convolutional Recurrent Neural Network (DCRNN)\mbox{\cite{li2017diffusion}}, Attention-Based Conv-LSTM Network (AT-Con-LSTM)\mbox{\cite{zheng2020hybrid}}, Bidirectional LSTM network\mbox{\cite{ma2021short}}, STFSA Convolutional Neural Network Gated Recurrent Unit (SCG)\mbox{\cite{ma2022novel}}, Adaptive Spatial-Temporal Fusion Graph Convolutional Network (AGFCRN)\mbox{\cite{li2022adaptive}} and Progressive Graph Convolutional Network (PGCN)\mbox{\cite{shin2024pgcn}}. Table \mbox{\ref{tab:Performance}} showcases the comparison of prediction accuracy (error rates) across different models using the MAE, MAPE, and RMSE indices. Notably, our proposed hybrid (AT-Bi-Conv-SE LSTM) model emerged as the overall best, consistently delivering exceptionally low MAE and RMSE rates of $7.49 \%$ and $11.03 \%$, respectively, for a 5-minute forecast. For the same prediction time, AGFCRN shows the highest MAE and RMSE rates of $15.10 \%$ and $17.81 \%$ respectively, making it the least effective for the same forecast duration. Other models, including DCRNN, AT-Con-LSTM, Bi-LSTM, SCG and PGCN, showed improved performances (descension of MAE and RMSE rates) over AGFCRN (least performing), with SCG being the superior model for shorter prediction times. Fig \mbox{~\ref{mse_rmsea}} affirms these findings, as we can visualize a reduction in the MAE and RMSE rates (improved model performance) moving from the least performing AGFCRN to the best performing SCG forecasting model for a 5-minute forecast. Similarly, from Table \mbox{\ref{tab:Performance}}, for a 60-minute forecast, AGFCRN remains the least efficient with the highest MAE and RMSE rates of $22.31 \%$ and $28.09 \%$ respectively, while our proposed model was the best-performing forecasting model with the least MAE and RMSE rates of $10.1 \%$ and $13.9 \%$ respectively (significantly reducing errors compared to AGFCRN and PGCN). Common to the behavior observed in Fig\mbox{~\ref{mse_rmsea}} for the 5-minute forecast, Fig\mbox{~\ref{mse_rmseb}} provides a visual illustration of the ascension in model performance for the forecasting algorithms moving from AGFCRN (the least performing algorithm), DCRNN, AT-Con-LSTM, Bi-LSTM, SCG, PGCN to our proposed hybrid model (best performing), in decreasing order of MAE and RMSE rates. It is essential to note, the trend of increasing MAE and RMSE rates with longer prediction horizons is consistent across all models as witnessed in both Tables \mbox{\ref{tab:Prediction}} and \mbox{\ref{tab:Performance}}. However, the SCG model, while excellent for short predictions (5 and 15 minutes), from Table \mbox{\ref{tab:Performance}}, falls short for longer horizons (30 and 60 minutes) compared to our proposed model. This indicating how reactionary the SCG model is, as well as underlining the superior capability of our proposed hybrid model in providing precise short-term traffic forecasts, essential for dynamic traffic management, incident response, and enhancing mobility, safety, and the overall efficiency of the transportation network.
    
\end{enumerate}

%%%%%%%%%%%%%%%%%%%%%%%%%%%%%%%%%%%%%%%%%%%%%%%%%%%

\label{subsec:results}

 \section*{Acknowledgment}
The work under this project is partly supported by the Department of Transportation Grants number 69A3552348327. Additionally, H. Mahmoud and A. Alsharif extend their appreciation to the Alabama Transportation Institute (ATI) for their support in conducting this research.

% The work under this project is supported by the National Science Foundation Grant number 2301553 and the NASA University Leadership Initiative (ULI) under grant number 80NSSC20M0161.

\section{Conclusion} \label{sec:Conclusions}
In this paper, we introduce a novel, lightweight traffic forecasting system specifically designed to safeguard the privacy of driver location information while delivering precise traffic flow predictions. Our approach employs IPFE technique to encrypt the location data of drivers to ensure the confidentiality of individual locations throughout the process. Each driver uses functional encryption to secure their location data, while TMCs employ a functional decryption key to compute route densities. Moreover, a hybrid Conv-LSTM and Bi-LSTM forecast model, which operates on encrypted route data, is utilized to predict near-future traffic densities. We conducted extensive simulations using real-world datasets to evaluate the efficacy of our system. The results highlight the efficiency, scalability, and low overhead in both computations and communications of system. Additionally, it demonstrated a reduction in data losses and improved traffic predictions, particularly at critical points of interest. Distinct from existing methods, our system ensures accurate forecasts without compromising privacy. Our hybrid Conv-LSTM and Bi-LSTM model, enhanced with a Squeeze-and-Excitation (SE) module, adeptly extracts spatial-temporal features from encrypted data, capturing intricate daily and weekly traffic patterns to boost prediction accuracy. Our findings confirm that our system adeptly meets the complex requirements of dynamic, real-time traffic management, setting a new benchmark in traffic forecasting with its ability to integrate seamlessly and operate reliably across various environments.

\newcolumntype{P}[1]{>{\centering\arraybackslash}p{#1}}

%\balance
\bibliographystyle{IEEEtran}
\bibliography{main}

% Generated by IEEEtran.bst, version: 1.14 (2015/08/26)
\begin{thebibliography}{10}
\providecommand{\url}[1]{#1}
\csname url@samestyle\endcsname
\providecommand{\newblock}{\relax}
\providecommand{\bibinfo}[2]{#2}
\providecommand{\BIBentrySTDinterwordspacing}{\spaceskip=0pt\relax}
\providecommand{\BIBentryALTinterwordstretchfactor}{4}
\providecommand{\BIBentryALTinterwordspacing}{\spaceskip=\fontdimen2\font plus
\BIBentryALTinterwordstretchfactor\fontdimen3\font minus
  \fontdimen4\font\relax}
\providecommand{\BIBforeignlanguage}[2]{{%
\expandafter\ifx\csname l@#1\endcsname\relax
\typeout{** WARNING: IEEEtran.bst: No hyphenation pattern has been}%
\typeout{** loaded for the language `#1'. Using the pattern for}%
\typeout{** the default language instead.}%
\else
\language=\csname l@#1\endcsname
\fi
#2}}
\providecommand{\BIBdecl}{\relax}
\BIBdecl

\bibitem{hansen2001determination}
I.~Hansen, ``Determination and evaluation of traffic congestion costs,''
  \emph{European Journal of Transport and Infrastructure Research}, vol.~1,
  no.~1, pp. 61--72, 2001.

\bibitem{inrix_2023}
\BIBentryALTinterwordspacing
Inrix, ``2022 inrix global traffic scorecard,'' January 2023. [Online].
  Available: \url{https://inrix.com/scorecard/}
\BIBentrySTDinterwordspacing

\bibitem{Sharma2022Awasthi}
S.~Sharma and S.~Awasthi, ``Introduction to intelligent transportation system:
  overview, classification based on physical architecture, and challenges,''
  \emph{Int. J. Sens. Netw.}, vol.~38, pp. 215--240, 2022.

\bibitem{Chhabra2021Krishna}
R.~Chhabra, C.~Krishna, and S.~Verma, ``A survey on state-of-the-art road
  surface monitoring techniques for intelligent transportation systems,''
  \emph{Int. J. Sens. Netw.}, vol.~37, pp. 81--99, 2021.

\bibitem{rabieh2016privacy}
K.~Rabieh, M.~M. Mahmoud, and M.~Younis, ``Privacy-preserving route reporting
  schemes for traffic management systems,'' \emph{IEEE Transactions on
  Vehicular Technology}, vol.~66, no.~3, pp. 2703--2713, 2016.

\bibitem{zhang2019privacy}
C.~Zhang, L.~Zhu, C.~Xu, X.~Du, and M.~Guizani, ``A privacy-preserving traffic
  monitoring scheme via vehicular crowdsourcing,'' \emph{Sensors}, vol.~19,
  no.~6, p. 1274, 2019.

\bibitem{zhang2017efficient}
Y.~Zhang, Q.~Pei, F.~Dai, and L.~Zhang, ``Efficient secure and
  privacy-preserving route reporting scheme for vanets,'' in \emph{Journal of
  Physics: Conference Series}, vol. 910, no.~1.\hskip 1em plus 0.5em minus
  0.4em\relax IOP Publishing, 2017, p. 012070.

\bibitem{lee2021vanet}
M.~Lee and T.~Atkison, ``Vanet applications: Past, present, and future,''
  \emph{Vehicular Communications}, vol.~28, p. 100310, 2021.

\bibitem{ma2020ridesharing}
J.~Ma, M.~Xu, Q.~Meng, and L.~Cheng, ``Ridesharing user equilibrium problem
  under od-based surge pricing strategy,'' \emph{Transportation Research Part
  B: Methodological}, vol. 134, pp. 1--24, 2020.

\bibitem{sun2020managing}
J.~Sun, J.~Wu, F.~Xiao, Y.~Tian, and X.~Xu, ``Managing bottleneck congestion
  with incentives,'' \emph{Transportation research part B: methodological},
  vol. 134, pp. 143--166, 2020.

\bibitem{huang2014deep}
W.~Huang, G.~Song, H.~Hong, and K.~Xie, ``Deep architecture for traffic flow
  prediction: deep belief networks with multitask learning,'' \emph{IEEE
  Transactions on Intelligent Transportation Systems}, vol.~15, no.~5, pp.
  2191--2201, 2014.

\bibitem{lv2014traffic}
Y.~Lv, Y.~Duan, W.~Kang, Z.~Li, and F.-Y. Wang, ``Traffic flow prediction with
  big data: a deep learning approach,'' \emph{IEEE Transactions on Intelligent
  Transportation Systems}, vol.~16, no.~2, pp. 865--873, 2014.

\bibitem{ma2017learning}
X.~Ma, Z.~Dai, Z.~He, J.~Ma, Y.~Wang, and Y.~Wang, ``Learning traffic as
  images: a deep convolutional neural network for large-scale transportation
  network speed prediction,'' \emph{Sensors}, vol.~17, no.~4, p. 818, 2017.

\bibitem{du2019deep}
B.~Du, H.~Peng, S.~Wang, M.~Z.~A. Bhuiyan, L.~Wang, Q.~Gong, L.~Liu, and J.~Li,
  ``Deep irregular convolutional residual lstm for urban traffic passenger
  flows prediction,'' \emph{IEEE Transactions on Intelligent Transportation
  Systems}, vol.~21, no.~3, pp. 972--985, 2019.

\bibitem{zheng2020hybrid}
H.~Zheng, F.~Lin, X.~Feng, and Y.~Chen, ``A hybrid deep learning model with
  attention-based conv-lstm networks for short-term traffic flow prediction,''
  \emph{IEEE Transactions on Intelligent Transportation Systems}, vol.~22,
  no.~11, pp. 6910--6920, 2020.

\bibitem{boneh2011functional}
D.~Boneh, A.~Sahai, and B.~Waters, ``Functional encryption: Definitions and
  challenges,'' in \emph{Theory of Cryptography: 8th Theory of Cryptography
  Conference, TCC 2011, Providence, RI, USA, March 28-30, 2011. Proceedings
  8}.\hskip 1em plus 0.5em minus 0.4em\relax Springer, 2011, pp. 253--273.

\bibitem{rabieh2015privacy}
K.~Rabieh, M.~M. Mahmoud, and M.~Younis, ``Privacy-preserving route reporting
  scheme for traffic management in vanets,'' in \emph{2015 IEEE International
  Conference on Communications (ICC)}.\hskip 1em plus 0.5em minus 0.4em\relax
  IEEE, 2015, pp. 7286--7291.

\bibitem{BBFog_Based}
J.~Zhang, H.~Fang, H.~Zhong, J.~Cui, and D.~He, ``Blockchain-assisted
  privacy-preserving traffic route management scheme for fog-based vehicular
  ad-hoc networks,'' \emph{IEEE Transactions on Network and Service
  Management}, vol.~20, no.~3, pp. 2854--2868, 2023.

\bibitem{EBCPA_9749919}
C.~Lin, X.~Huang, and D.~He, ``Ebcpa: Efficient blockchain-based conditional
  privacy-preserving authentication for vanets,'' \emph{IEEE Transactions on
  Dependable and Secure Computing}, vol.~20, no.~3, pp. 1818--1832, 2023.

\bibitem{PP_CA9928430}
Y.~Qi, J.~Wu, A.~K. Bashir, X.~Lin, W.~Yang, and M.~D. Alshehri,
  ``Privacy-preserving cross-area traffic forecasting in its: A transferable
  spatial-temporal graph neural network approach,'' \emph{IEEE Transactions on
  Intelligent Transportation Systems}, vol.~24, no.~12, pp. 15\,499--15\,512,
  2023.

\bibitem{LSM_TC10324334}
D.~Wang, W.~Li, and J.~Pan, ``Large-scale mixed traffic control using dynamic
  vehicle routing and privacy-preserving crowdsourcing,'' \emph{IEEE Internet
  of Things Journal}, vol.~11, no.~2, pp. 1981--1989, 2024.

\bibitem{calandriello2007efficient}
G.~Calandriello, P.~Papadimitratos, J.-P. Hubaux, and A.~Lioy, ``Efficient and
  robust pseudonymous authentication in vanet,'' in \emph{Proceedings of the
  fourth ACM international workshop on Vehicular ad hoc networks}, 2007, pp.
  19--28.

\bibitem{xia2022short}
M.~Xia, D.~Jin, and J.~Chen, ``Short-term traffic flow prediction based on
  graph convolutional networks and federated learning,'' \emph{IEEE
  Transactions on Intelligent Transportation Systems}, vol.~24, no.~1, pp.
  1191--1203, 2022.

\bibitem{levin1980forecasting}
M.~Levin and Y.-D. Tsao, ``On forecasting freeway occupancies and volumes
  (abridgment),'' \emph{Transportation Research Record}, no. 773, 1980.

\bibitem{hamed1995short}
M.~M. Hamed, H.~R. Al-Masaeid, and Z.~M.~B. Said, ``Short-term prediction of
  traffic volume in urban arterials,'' \emph{Journal of Transportation
  Engineering}, vol. 121, no.~3, pp. 249--254, 1995.

\bibitem{van1996combining}
M.~Van Der~Voort, M.~Dougherty, and S.~Watson, ``Combining kohonen maps with
  arima time series models to forecast traffic flow,'' \emph{Transportation
  Research Part C: Emerging Technologies}, vol.~4, no.~5, pp. 307--318, 1996.

\bibitem{lee1999application}
S.~Lee and D.~B. Fambro, ``Application of subset autoregressive integrated
  moving average model for short-term freeway traffic volume forecasting,''
  \emph{Transportation research record}, vol. 1678, no.~1, pp. 179--188, 1999.

\bibitem{williams2003modeling}
B.~M. Williams and L.~A. Hoel, ``Modeling and forecasting vehicular traffic
  flow as a seasonal arima process: Theoretical basis and empirical results,''
  \emph{Journal of transportation engineering}, vol. 129, no.~6, pp. 664--672,
  2003.

\bibitem{feng2018adaptive}
X.~Feng, X.~Ling, H.~Zheng, Z.~Chen, and Y.~Xu, ``Adaptive multi-kernel svm
  with spatial--temporal correlation for short-term traffic flow prediction,''
  \emph{IEEE Transactions on Intelligent Transportation Systems}, vol.~20,
  no.~6, pp. 2001--2013, 2018.

\bibitem{10145923}
Y.~Miao, X.~Bai, Y.~Cao, Y.~Liu, F.~Dai, F.~Wang, L.~Qi, and W.~Dou, ``A novel
  short-term traffic prediction model based on svd and arima with blockchain in
  industrial internet of things,'' \emph{IEEE Internet of Things Journal},
  vol.~10, no.~24, pp. 21\,217--21\,226, 2023.

\bibitem{tan2009aggregation}
M.-C. Tan, S.~C. Wong, J.-M. Xu, Z.-R. Guan, and P.~Zhang, ``An aggregation
  approach to short-term traffic flow prediction,'' \emph{IEEE Transactions on
  Intelligent Transportation Systems}, vol.~10, no.~1, pp. 60--69, 2009.

\bibitem{SPTGCN9945663}
Y.~Zhao, Y.~Lin, H.~Wen, T.~Wei, X.~Jin, and H.~Wan, ``Spatial-temporal
  position-aware graph convolution networks for traffic flow forecasting,''
  \emph{IEEE Transactions on Intelligent Transportation Systems}, vol.~24,
  no.~8, pp. 8650--8666, 2023.

\bibitem{Inception10032279}
Y.~Wang, C.~Jing, W.~Huang, S.~Jin, and X.~Lv, ``Adaptive spatiotemporal
  inceptionnet for traffic flow forecasting,'' \emph{IEEE Transactions on
  Intelligent Transportation Systems}, vol.~24, no.~4, pp. 3882--3907, 2023.

\bibitem{CNNGRU9701439}
C.~Ma, Y.~Zhao, G.~Dai, X.~Xu, and S.-C. Wong, ``A novel stfsa-cnn-gru hybrid
  model for short-term traffic speed prediction,'' \emph{IEEE Transactions on
  Intelligent Transportation Systems}, vol.~24, no.~4, pp. 3728--3737, 2023.

\bibitem{ma2022novel}
------, ``A novel stfsa-cnn-gru hybrid model for short-term traffic speed
  prediction,'' \emph{IEEE Transactions on Intelligent Transportation Systems},
  vol.~24, no.~4, pp. 3728--3737, 2022.

\bibitem{tian2015predicting}
Y.~Tian and L.~Pan, ``Predicting short-term traffic flow by long short-term
  memory recurrent neural network,'' in \emph{2015 IEEE international
  conference on smart city/SocialCom/SustainCom (SmartCity)}.\hskip 1em plus
  0.5em minus 0.4em\relax IEEE, 2015, pp. 153--158.

\bibitem{zhaowei2020short}
Q.~Zhaowei, L.~Haitao, L.~Zhihui, and Z.~Tao, ``Short-term traffic flow
  forecasting method with mb-lstm hybrid network,'' \emph{IEEE Transactions on
  Intelligent Transportation Systems}, vol.~23, no.~1, pp. 225--235, 2020.

\bibitem{ma2021short}
C.~Ma, G.~Dai, and J.~Zhou, ``Short-term traffic flow prediction for urban road
  sections based on time series analysis and lstm\_bilstm method,'' \emph{IEEE
  Transactions on Intelligent Transportation Systems}, vol.~23, no.~6, pp.
  5615--5624, 2021.

\bibitem{cheng9345387}
Z.~Cheng, J.~Lu, H.~Zhou, Y.~Zhang, and L.~Zhang, ``Short-term traffic flow
  prediction: An integrated method of econometrics and hybrid deep learning,''
  \emph{IEEE Transactions on Intelligent Transportation Systems}, vol.~23,
  no.~6, pp. 5231--5244, 2022.

\bibitem{10.1007/978-3-642-19571-6_16}
D.~Boneh, A.~Sahai, and B.~Waters, ``Functional encryption: Definitions and
  challenges,'' \emph{in Theory of Cryptography: Springer Berlin Heidelberg},
  pp. 253--273, Mar. 2011.

\bibitem{10.1007/978-3-662-53015-3_12}
S.~Agrawal, B.~Libert, and D.~Stehl{\'e}, ``Fully secure functional encryption
  for inner products, from standard assumptions,'' \emph{in Advances in
  Cryptology -- CRYPTO 2016: Springer Berlin Heidelberg}, pp. 333--362, Jul.
  2016.

\bibitem{10.1007/978-3-662-46447-2_33}
M.~Abdalla, F.~Bourse, A.~De~Caro, and D.~Pointcheval, ``Simple functional
  encryption schemes for inner products,'' \emph{in Public-Key Cryptography --
  PKC 2015: Springer Berlin Heidelberg}, pp. 733--751, Mar. 2015.

\bibitem{10.1007/978-3-319-63688-7_3}
C.~Baltico, D.~Catalano, D.~Fiore, and R.~Gay, ``Practical functional
  encryption for quadratic functions with applications to predicate
  encryption,'' \emph{in Advances in Cryptology -- CRYPTO 2017, Cham: Springer
  International Publishing}, pp. 67--98, Jul. 2017.

\bibitem{hu2018squeeze}
J.~Hu, L.~Shen, and G.~Sun, ``Squeeze-and-excitation networks,'' in
  \emph{Proceedings of the IEEE conference on computer vision and pattern
  recognition}, 2018, pp. 7132--7141.

\bibitem{sweeney2002k}
L.~Sweeney, ``k-anonymity: A model for protecting privacy,''
  \emph{International journal of uncertainty, fuzziness and knowledge-based
  systems}, vol.~10, no.~05, pp. 557--570, 2002.

\bibitem{escala2017algebraic}
A.~Escala, G.~Herold, E.~Kiltz, C.~R{\`a}fols, and J.~Villar, ``An algebraic
  framework for diffie--hellman assumptions,'' \emph{Journal of cryptology},
  vol.~30, pp. 242--288, 2017.

\bibitem{10.1007/3-540-69053-0_18}
V.~Shoup, ``Lower bounds for discrete logarithms and related problems,''
  \emph{in Advances in Cryptology --- EUROCRYPT '97: Springer Berlin
  Heidelberg}, pp. 256--266, Jul. 1997.

\bibitem{OpenStreetMap}
\BIBentryALTinterwordspacing
{OpenStreetMap}. \url{https://www.openstreetmap.org}. Accessed on 23 October
  2023. [Online]. Available: \url{https://www.openstreetmap.org}
\BIBentrySTDinterwordspacing

\bibitem{ITSBerlin2015}
\BIBentryALTinterwordspacing
{Institute of Transportation Systems at Berlin}. (2015) {SUMO - Simulation of
  Urban MObility}. Accessed on 23 October 2023. [Online]. Available:
  \url{"http://www.dlr.de/ts/en/desktopdefault.aspx/tabid-9883/16931read-41000/"}
\BIBentrySTDinterwordspacing

\bibitem{PeMS}
\BIBentryALTinterwordspacing
{California Department of Transportation}. (Year the page was last updated, if
  available) {Performance Measurement System (PeMS) Data Source}.
  \url{https://dot.ca.gov/programs/traffic-operations/mpr/pems-source}.
  Accessed on 23 October 2023. [Online]. Available:
  \url{https://dot.ca.gov/programs/traffic-operations/mpr/pems-source}
\BIBentrySTDinterwordspacing

\bibitem{li2017diffusion}
Y.~Li, R.~Yu, C.~Shahabi, and Y.~Liu, ``Diffusion convolutional recurrent
  neural network: Data-driven traffic forecasting,'' \emph{arXiv preprint
  arXiv:1707.01926}, 2017.

\bibitem{li2022adaptive}
S.~Li, L.~Ge, Y.~Lin, and B.~Zeng, ``Adaptive spatial-temporal fusion graph
  convolutional networks for traffic flow forecasting,'' in \emph{2022
  International Joint Conference on Neural Networks (IJCNN)}.\hskip 1em plus
  0.5em minus 0.4em\relax IEEE, 2022, pp. 1--8.

\bibitem{shin2024pgcn}
Y.~Shin and Y.~Yoon, ``Pgcn: Progressive graph convolutional networks for
  spatial--temporal traffic forecasting,'' \emph{IEEE Transactions on
  Intelligent Transportation Systems}, 2024.

\bibitem{girija2016tensorflow}
S.~S. Girija, ``Tensorflow: Large-scale machine learning on heterogeneous
  distributed systems,'' \emph{Software available from tensorflow. org},
  vol.~39, no.~9, 2016.

\end{thebibliography}

\newpage
\section*{Biographies}

%%%%%%%%%%%%%%%%%%%%%%%%%%%%%%%%%%%%%%%%%%

\begin{IEEEbiography}[{\includegraphics[width=1in,height=2.4in,clip,keepaspectratio]{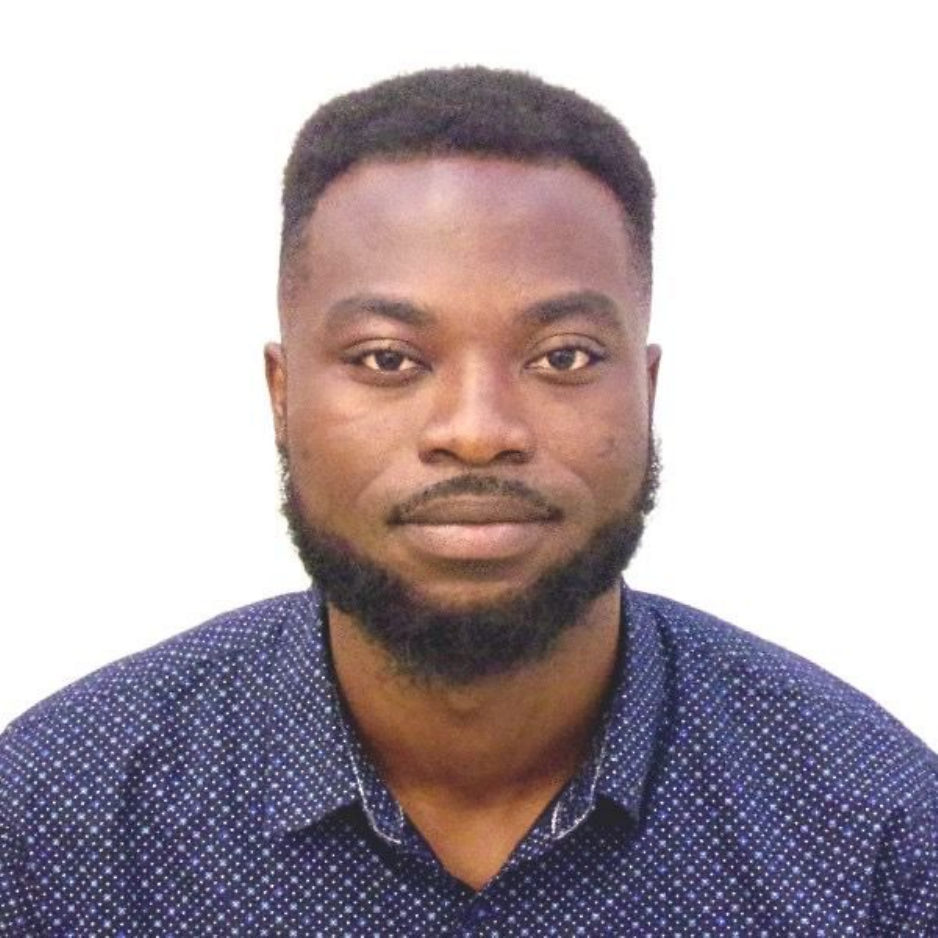}}]{Isaac Adom} received his Electrical and Electronic Engineering Bachelor of Science (B.Sc.) degree from Kwame Nkrumah University of Science and Technology (KNUST), Kumasi, Ghana in 2014. He is currently pursuing his Doctorate (Ph.D.) degree with the Department of Electrical and Computer Engineering, North Carolina A\&T State University, USA, where he holds a graduate research assistant position with research interests that include eXplainable Artificial Intelligence (XAI) for autonomous systems, cybersecurity and privacy for vehicular ad hoc networks, federated learning, smart grids, unmanned aerial vehicles, and other machine learning applications

.
\end{IEEEbiography}

  \begin{IEEEbiography}[{\includegraphics[width=1in,height=2.4in,clip,keepaspectratio]{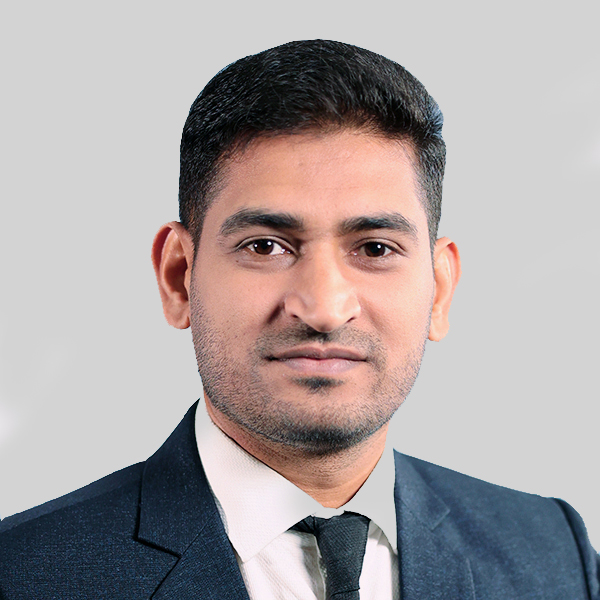}}]{Mohamed Iqbal Hossain}
received a Bachelor of Science in Electrical and Electronic Engineering from Rajshahi University of Engineering \& Technology, Rajshahi, Bangladesh 2012. He completed his Master of Science in Electrical Engineering from North Carolina A\&T State University, USA, in 2023. He worked as a Research Assistant during his Master's program funded by Lockheed Martin. He is currently an Associate Distribution Engineer at Leidos. His research interests include Cyber Security, Power Systems, Smart Grids, Control Systems, and Artificial Intelligence.
\end{IEEEbiography}
%%%%%%%%%%%%%%%%%%%%%%%%%%%%%%%%%%%%%%%%%%
\begin{IEEEbiography}
[{\includegraphics[width=1in,height=1.25in,clip,keepaspectratio]{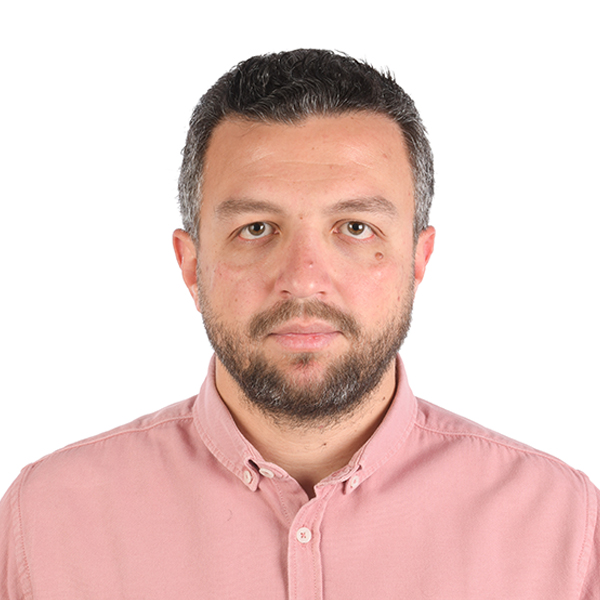}}]{Hassan Mahmoud} (S'23) received the B.Sc. degree in Electronic and Electrical Communication from Faculty of Engineering, Helwan university, Cairo, Egypt and M.Sc. degree in Electronic and Communication Engineering from the College of Engineering and Technology, The Arab Academy for Science, Technology and Maritime Transport, Cairo, Egypt in 2009 and 2021 respectively. He is currently working towards the Ph.D. degree at Department of Computer Science, College of Engineering, The University of Alabama, Tuscaloosa, AL, USA, where he is also a Graduate Research Assistant. His research interests include security and privacy in smart grid, vehicular ad hoc networks, and intelligent transportation systems.
\end{IEEEbiography}

%%%%%%%%%%%%%%%%%%%%%%%%%%%%%%%%%%%%%%%%%%

%%%%%%%%%%%%%%%%%%%%%%%%%%%%%%%%%%%%%%%%%%
\begin{IEEEbiography}
[{\includegraphics[width=1in,height=1.25in,clip,keepaspectratio]{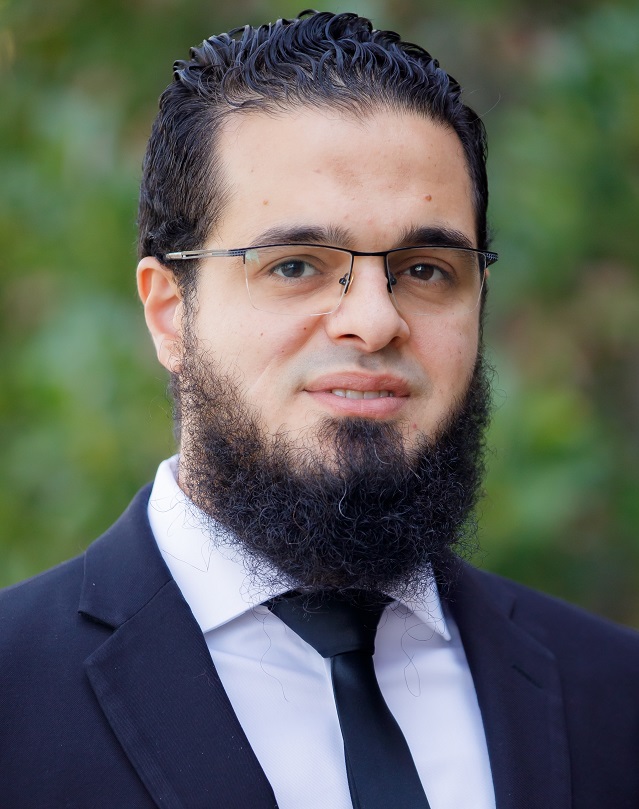}}]{Ahmad Alsharif} (M'18, SM'21) received the B.Sc. and M.Sc. degrees (Hons.) in electrical engineering from Benha University, Cairo, Egypt, in 2009 and 2015, respectively, and the Ph.D. degree in electrical and computer engineering from Tennessee Tech University, Cookeville, TN, USA, in May 2019. He is an Assistant Professor at The University of Alabama, Tuscaloosa, AL, USA. He also holds the position of an Assistant Professor with the Faculty of Engineering, Shoubra, Benha University. His current research interests include applied cryptography, secure protocol design, IoT security, cyber–physical systems security, digital forensics, secure blockchain applications, and machine learning applications in cybersecurity. Dr. Alsharif was awarded the U.S. National Science Foundation Research Initiation Initiative Grant (NSF CRII) in 2022. He received the Young Innovator Award from the Egyptian Industrial Modernisation Center in 2009.
\end{IEEEbiography}

%%%%%%%%%%%%%%%%%%%%%%%%%%%%%%%%%%%%%%%%%%
\begin{IEEEbiography}[{\includegraphics[width=1in,height=1.25in,clip,keepaspectratio]{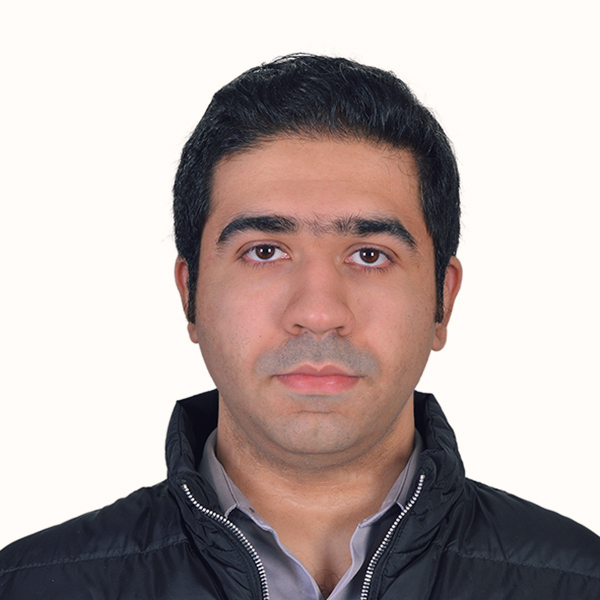}}]{Dr. Mahmoud Nabil} (Member, IEEE) Dr. Mahmoud received his Bachelor of Science (BS) and Master of Science (MS) degrees, with honors, in computer engineering from Cairo University, Egypt, in 2012 and 2016, respectively. He completed his Ph.D. in electrical and computer engineering from Tennessee Tech University, Cookeville, TN, USA, in August 2019. He is an assistant professor at the Department of Electrical and Computer Engineering at North Carolina A \& T State University. Dr. Mahmoud is an accomplished researcher and has authored and co-authored numerous publications in prestigious venues. His research has been published in renowned journals such as IEEE Internet of Things, IEEE Transactions of Dependable and Secure Computing, IEEE Transactions on Human-Machine Systems, and IEEE Transactions of Mobile Computing. He has also contributed to leading conferences including the International Conference on Communication, International Conference on Pattern Recognition, and International Conference on Wireless Communication. With diverse research interests, Dr. Mahmoud's areas of expertise include security and privacy in unmanned aerial systems, smart grids, machine learning applications, vehicular Ad Hoc networks, and blockchain applications. He has received significant funding for his research projects from esteemed national agencies and organizations, including the National Science Foundation (NSF), Department of Transportation (DOT), Air Force Research Laboratory (AFRL), NASA, Intel, Cisco, and Lockheed Martin.
\end{IEEEbiography}

%%%%%%%%%%%%%%%%%%%%%%%%%%%%%%%%%%%%%%%%%%
%%%%%%%%%%%%%%%%%%%%%%%%%%%%%%%%%%%%%%%%%%
\begin{IEEEbiography}[{\includegraphics[width=1in,height=1.25in,clip,keepaspectratio]{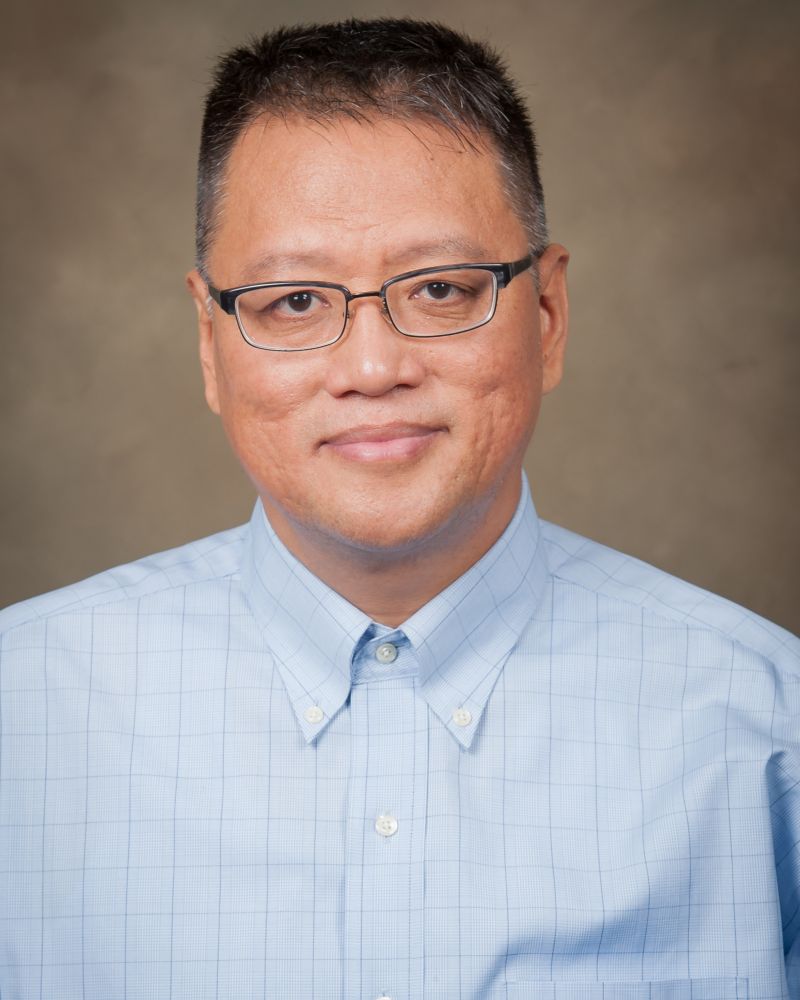}}]{Yang Xiao}  (Fellow, IEEE) received the B.S. and M.S. degrees in computational mathematics from Jilin University, Changchun, China, in 1989 and 1991, respectively, and the M.S. and Ph.D. degrees in computer science and engineering from Wright State University, Dayton, OH, USA, in 2000 and 2001, respectively. He is a Full Professor with the Department of Computer Science, The University of Alabama, Tuscaloosa, AL, USA. He directed over 20 doctoral dissertations and supervised over 20 M.S. theses/projects. He has published over 300 Science Citation Index (SCI)-indexed journal papers (including over 60 IEEE/ACM TRANSACTIONS) and 300 Engineering Index (EI)-indexed refereed conference papers and book chapters related to these research areas. His research interests include cyber–physical systems, the Internet of Things, security, wireless networks, smart grid, and telemedicine. Prof. Xiao was a Voting Member of the IEEE 802.11 Working Group from 2001 to 2004, involving the IEEE 802.11 (Wi-Fi) standardization work. He is an IEEE Fellow, an IET Fellow, and an AAIA Fellow. He served as a Guest Editor over 35 times for different international journals, including the IEEE Journal on Selected Areas in Communications (JSAC) in 2022-2023, IEEE TRANSACTIONS ON NETWORK SCIENCE AND ENGINEERING in 2021, IEEE TRANSACTIONS ON GREEN COMMUNICATIONS AND NETWORKING in 2021, IEEE Network in 2007, IEEE WIRELESS COMMUNICATIONS in 2006 and 2021, IEEE Communications Standards Magazine in 2021, and Mobile Networks and Applications (MONET) (ACM/Springer) in 2008. He also serves as the Editor-in-Chief of Cyber-Physical Systems Journal, International Journal of Sensor Networks (IJSNet), and International Journal of Security and Networks (IJSN). He has been serving as an Editorial Board Member or an Associate Editor for 20 international journals, including the IEEE Transactions on Network Science and Engineering (TNSE) since 2022, IEEE TRANSACTIONS ON CYBERNETICS since 2020, IEEE TRANSACTIONS ON SYSTEMS, MAN, AND CYBERNETICS: SYSTEMS from 2014 to 2015, IEEE TRANSACTIONS ON VEHICULAR TECHNOLOGY from 2007 to 2009, and IEEE COMMUNICATIONS SURVEYS AND TUTORIALS from 2007 to 2014. He serves/served as a Member of the Technical Program Committee for more than 300 conferences. He received the IEEE Transactions on Network Science and Engineering Excellent Editor Award 2022.
\end{IEEEbiography} 

\end{document}